\documentclass[aps,notitlepage,nofootinbib,superscriptaddress,twocolumn,longbibliography,10pt]{revtex4-1}

\usepackage[normalem]{ulem}

\newcommand{\beq}{\begin{eqnarray}}
\newcommand{\eeq}{\end{eqnarray}}
\newcommand{\beqq}{\begin{eqnarray*}}
\newcommand{\eeqq}{\end{eqnarray*}}

\newcommand{\be}{\begin{equation}}
\newcommand{\ee}{\end{equation}}
\newcommand{\barr}{\begin{array}}
\newcommand{\earr}{\end{array}}

\newcommand{\tr}{\text{tr}}

\newcommand{\ztyp}{Z^\textrm{typ}}

\newcommand{\f}{\frac}

\newcommand{\lf}{\left(}
\newcommand{\ri}{\right)}

\newcommand{\bra}[1]{\< #1 \right|}
\newcommand{\ket}[1]{\left| #1 \>}

\usepackage{amsmath}
\usepackage{amsthm, amssymb}

\usepackage[colorlinks=true,citecolor=blue,linkcolor=blue]{hyperref}
\usepackage{graphicx}
\usepackage{dcolumn}
\usepackage{bm}
\usepackage{color}
\usepackage{tabularx}
\usepackage{comment}
\usepackage{float}
\usepackage{amsmath}
\usepackage{gensymb}
\usepackage{mathtools}
\usepackage{tikz-cd}
\usepackage{adjustbox}
\usepackage{braket}
\usepackage{dsfont}

\begin{document}

\begin{titlepage}

\widetext

\title{Measurement-induced phase transitions on dynamical quantum trees}

\author{Xiaozhou Feng}
\affiliation{Department of Physics, The Ohio State University, Columbus,
Ohio 43202, USA}

\author{Brian Skinner}
\affiliation{Department of Physics, The Ohio State University, Columbus,
Ohio 43202, USA}

\author{Adam Nahum}
\affiliation{Laboratoire de Physique de l'\'Ecole Normale Sup\'erieure, ENS, Universit\'e PSL, CNRS, Sorbonne Universit\'e, Universit\'e Paris-Diderot, Sorbonne Paris Cit\'e, Paris, France.}

\setcounter{equation}{0}
\setcounter{figure}{0}
\setcounter{table}{0}

\makeatletter
\renewcommand{\theequation}{S\arabic{equation}}
\renewcommand{\thefigure}{S\arabic{figure}}
\renewcommand{\thetable}{S\Roman{table}}
\renewcommand{\bibnumfmt}[1]{[S#1]}
\renewcommand{\citenumfont}[1]{S#1}

\date{\today}

\begin{abstract}
Monitored many-body systems fall broadly into two dynamical phases, ``entangling'' or ``disentangling'', separated by a transition as a function of the rate at which measurements are made on the system. Producing an analytical theory of this measurement-induced transition is an outstanding challenge. Recent work made progress in the context of tree tensor networks, which can be related to all-to-all quantum circuit dynamics with forced (postselected) measurement outcomes. So far, however, there are no exact solutions for dynamics of spin-1/2 degrees of freedom (qubits) with ``real'' measurements, whose  outcome probabilities are sampled according to the Born rule. Here we define dynamical processes for qubits, with real measurements, that have a tree-like spacetime interaction graph, either collapsing or expanding the system as a function of time. The former case yields an exactly solvable measurement transition. We explore these processes analytically and numerically, exploiting the recursive structure of the tree. We compare the case of ``real'' measurements with the case of ``forced'' measurements. Both cases show a transition at a nontrivial value of the measurement strength, with the real measurement case exhibiting a smaller entangling phase. Both exhibit exponential scaling of the entanglement near the transition, but they differ in the value of a critical exponent. An intriguing difference between the two cases is that the real measurement case lies at the  boundary between two distinct types of critical scaling. On the basis of our results we  propose a protocol for realizing a measurement phase transition experimentally via an expansion process.
\end{abstract}

\pacs{}

\maketitle

\draft

\vspace{2mm}

\end{titlepage}

\section{Introduction}
\label{sec:intro}

If the unitary time evolution of a many-body quantum system is punctuated by measurements, made at a finite rate per local degree of freedom,  the resulting dynamics can fall into one of two broad classes known as the ``entangling'' and ``disentangling'' dynamical phases, or the ``weak monitoring'' and ``strong monitoring'' phases. Separating these two phases is a measurement-induced phase transition (MPT), which can be crossed 
by increasing the rate at which measurements are performed 
\cite{Skinner_Measurement_2019,Li_Quantum_2018,Chan_Unitary-projective_2019,Li_Measurement-driven_2019,Szyniszewski_Entanglement_2019,Choi_Quantum_2020,Gullans_Dynamical_2020,Bao_Theory_2020,Jian_Measurement-induced_2020,Li_Conformal_2021,Zabalo_Critical_2020,Gullans_Scalable_2020,Tang_Measurement-induced_2020,Fuji_Measurement-induced_2020,Lunt_Measurement-induced_2020,Szyniszewski_Universality_2020,Turkeshi_Measurement-induced_2020,Van_Entanglement_2021,Fan_Self-organized_2021,Li_Statistical_2021,Shtanko_Classical_2020,Nahum_Measurement_2021,Vijay_Measurement_2020}.

In the original identification of the MPT \cite{Skinner_Measurement_2019, Li_Quantum_2018}, the transition was associated with the scaling properties of the entanglement entropy of a pure state (volume-law versus area-law). Subsequent work made clear that there are alternative useful formulations of the MPT. 
In particular, one can associate the MPT with a sharp change in the rate at which an initially mixed state is transformed into a pure state by the monitored dynamics \cite{Gullans_Dynamical_2020}.
In the disentangling phase the characteristic timescale for purification is only order-one, 
while in the entangling phase 
purification takes a time that is exponential in the system size. 
This purification timescale  is also the timescale up to which quantum information from the initial state is retained in the final state \cite{Nahum_Measurement_2021,Choi_Quantum_2020,fidkowski2021dynamical,Li_Statistical_2021}, 
and is therefore related to questions like ``do I need to know the initial state in order to deduce the final state, or can I deduce it from the measurement outcomes alone?''.

While numerical studies have characterized the MPT in a wide variety of settings (see Refs.~\cite{reviewentanglementdynamicsinhybridquantumcircuits,reviewrandomquantumcircuits2022} for reviews),
so far its critical properties have largely evaded an exact analytical treatment (except in the limit of infinite local Hilbert space dimension, where the phase transition reduces to a problem of the classical geometry of the circuit \cite{Skinner_Measurement_2019, Nahum_Quantum_2017}). 
There is an active effort to describe the MPT and other models with related mathematical structure using effective statistical mechanical lattice models
\cite{Bao_Theory_2020,Jian_Measurement-induced_2020,Nahum_Measurement_2021,Vasseur_Entanglement_2019,Hayden_Holographic_2016,Zhou_Emergent_2019, Nahum_Operator_2018,li2021statistical,barratt2021field, nahum2020entanglement,
sang2021measurement,
Lavasani_Measurement_2021,
Ippoliti_Entanglement_2021,
Lang_Entanglement_2020,
Turkeshi_Entanglement_2023,
Le_Gal_Volume_2022}
or effective ``Landau-Ginsburg-like'' field theories \cite{Vasseur_Entanglement_2019,Nahum_Measurement_2021}.
But so far there is no theory that can reproduce the numerically observed critical measurement rate nor the critical exponents of the MPT in a low-dimensional system of qubits.

This challenge has motivated
simplifying the spatial structure of the problem.
Recent work  examined the MPT in all-to-all coupled circuits \cite{Gullans_Dynamical_2020, Vijay_Measurement_2020, Nahum_Measurement_2021}. In these models unitary gates are applied (at random, with equal rates) between any pair of spins/qubits, and these unitaries are  interspersed with a finite rate of single-spin projective measurements. 
Such models take the spin system to a mean-field-like limit in which all pairs of spins can interact. 
The hope is that this limit will enable an analytical approach that is not available in finite-dimensional systems. 

Recent progress in this direction has made use of the fact that  the quantum circuit for a large all-to-all-coupled system has a locally tree-like structure \cite{Nahum_Measurement_2021},
leading to the conjecture that the MPT in the all-to-all circuit coincides with an entanglement transition in an associated ensemble of tree tensor networks.
This correspondence enables exact results in the case where the measurement outcomes are ``forced'', as described below.

In order to characterize the entanglement transition on a tree tensor network, one can define an ``order parameter'' $Z_k$ (defined more precisely below) which quantifies the degree of entanglement between the root and the leaves of a tree with $k$ generations. In the weak monitoring phase, this order parameter remains nonzero when the tree's size diverges.
Importantly, the tree structure allows one to solve for the behavior of $Z_k$ as $k\rightarrow \infty$ using a recursion relation,  giving analytic results for the critical measurement rate and the critical vanishing of $Z_k$ near the transition \cite{Nahum_Measurement_2021}. These results generalize to entanglement transitions in a large class of tree tensor networks. 
These developments, as well as the results in \cite{Lopez-Piqueres_Mean-field_2020,deterministictreeinprep},
suggest that tree tensor networks \cite{Shi_Classical_2006,Tagliacozzo_Simulation_2009,Murg_Simulating_2010,Silvi_Homogeneous_2010,Li_Efficient_2012,Murg_Tree_2015,Vidal_Entanglement_2007,Swingle_Entanglement_2012,Lopez-Piqueres_Mean-field_2020,Nakatani_Efficient_2013} are a useful tool for developing a mean-field theory for measurement and entanglement transitions.

Despite this progress, we  do not yet have an exact solution for an MPT for qubits with \textit{true} measurements, even in the context of quantum trees.
In Ref.~\onlinecite{Nahum_Measurement_2021},
a reduction to a tractable tree tensor network was obtained  only for a circuit  
where measurement outcomes were predetermined, instead of being sampled with Born's rule. Physically, the former situation corresponds to a protocol where one runs the dynamics many times and then discards all realizations except those for which  the measurements produce a desired sequence of outcomes (e.g., all spins are measured in the $\uparrow$ state). This protocol was dubbed the ``forced measurement'' case, and the resulting transition is called the ``forced-measurement-induced phase transition'' (FMPT). 
A key question, then, is whether the approach of Ref.~\onlinecite{Nahum_Measurement_2021} can be extended to describe \emph{real} measurements, for which the measurement outcomes
are sampled with the nontrivial, state-dependent probability given by Born's rule.
More generally, one can ask how the MPT and FMPT differ in various settings. Do they have different critical points and/or different critical exponents? It would be valuable to have a model in which exact results can be obtained for a version of the MPT.

\begin{figure}
    \centering
    \includegraphics[width = 1.0\columnwidth]{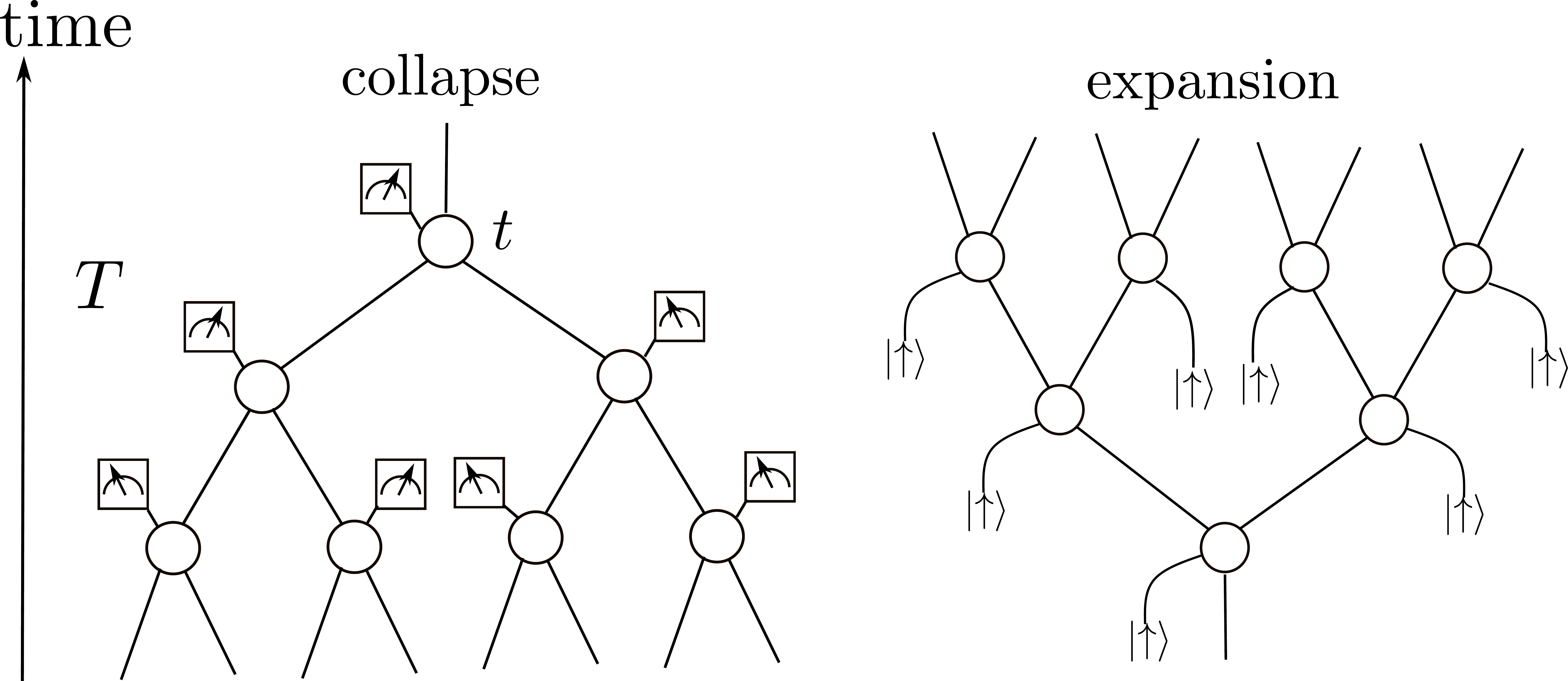}
    \caption{Schematic illustrations of the collapse process (Left) and expansion processes (Right).
    Each node of the tree represents an interaction between two qubits, preceded/followed by weak measurements.
    Viewing the tree as a tensor network, the symbol $t$ denotes a local three-legged tensor and $T$ denotes the full network.  (These tensors are defined in the  text.)  }
    \label{fig:collapse_expansion}
\end{figure}

In this paper we define a different kind of many-body dynamics,
the ``collapse process'', which has
a solvable phase transition for real measurements as well as for forced measurements. This solvability allows us to examine the difference between real and forced measurements. 
We also define a closely-related ``expansion process''.
Cartoons of these two processes are shown in Fig.~\ref{fig:collapse_expansion}.
A potentially useful feature of dynamical trees is that they exhibit measurement phase transitions that are experimentally accessible without the need for postselection
\cite{Gullans_Scalable_2020,Noel_Measurement_2022,Ippoliti_postselection_2021,Koh_Experimental_2022,Li_Cross_2022,garratt2022measurements}
on measurement outcomes, because the  efficient contractability of tree tensor networks can be exploited to classically process experimental measurement outcomes.

The dynamics we discuss are not conventional quantum circuit dynamics with a fixed number of qubits. 
Instead we define dynamical protocols whose spacetime interaction graph resembles a tree, with the number of active qubits either decreasing or increasing in time. In the ``quantum collapse process'' (Fig.~\ref{fig:collapse_expansion}, left) we begin with ${2^k\gg 1}$ qubits, and by measuring and discarding half of the qubits 
at each timestep, we are left with a single qubit at time $t=k$. 
Interactions and weak measurements also take place during each timestep.
In the ``expansion process'' (Fig.~\ref{fig:collapse_expansion}, right) we start with a single qubit and (by recruiting further spins) attempt to scramble its state into a many-body state of $2^t$ qubits.  In both cases, we ask whether a maximally mixed initial state is purified by the monitored dynamics (the strong monitoring phase) or not (the weak monitoring phase). 
The purity of the state is quantified by an entropy that varies between 0 and 1 bit (since the root of the tree is a single spin); this entropy also represents the entanglement entropy between the root and the leaves.
The entanglement transition is induced by varying the strength of the weak measurements that occur in every timestep.
We focus mainly on the collapse process, where we give exact results for both MPT and FMPT. 
(Our results imply exact results for the expansion process FMPT, but not the expansion process MPT, as we discuss in more detail below.)

Let us contrast the ``dynamical'' trees studied here with the trees studied in  
Ref.~\cite{Nahum_Measurement_2021}.
In the present work, the  ``generation'' coordinate of the tree is simply proportional to the physical time coordinate $t$.
By contrast, in  Ref.~\cite{Nahum_Measurement_2021} trees arose  more indirectly, starting from circuit dynamics with a fixed number of qubits:  in that case the  ``generation'' coordinate of the tree was \textit{not} equivalent to physical time.\footnote{In the previous case the generation coordinate represented the graph distance from an arbitrarily chosen bond of the circuit (i.e. an arbitrary point on the worldline of some spin): see Fig.~4 of~Ref.~\onlinecite{Nahum_Measurement_2021}.}
Ultimately, in both cases we must deal with a random ensemble of tree tensor networks. However, the fact that we consider real measurements here means the probability distribution for this ensemble of trees has nontrivial correlations between different nodes.

Crucially, the collapse model we introduce maintains
a unitary invariance property for the distribution of the output states,
even for the case of real measurements:
as a result of the Haar-randomness of the interaction unitaries, there is no preferred local basis.
This invariance allows us to
locate the transition analytically by studying a recursive map  \cite{Nahum_Measurement_2021}. The invariance also constrains the universality class of the phase transition \cite{Nahum_Measurement_2021, deterministictreeinprep}. Interestingly, the critical scaling forms are similar to  those describing branching random walks with a moving absorbing wall \cite{derrida_survival_2007}: the reason is that both problems are closely connected to traveling wave equations \cite{Derrida_Polymers_1988}.

\begin{figure}[tb]
    \centering
    \includegraphics[width= 1.0\columnwidth]{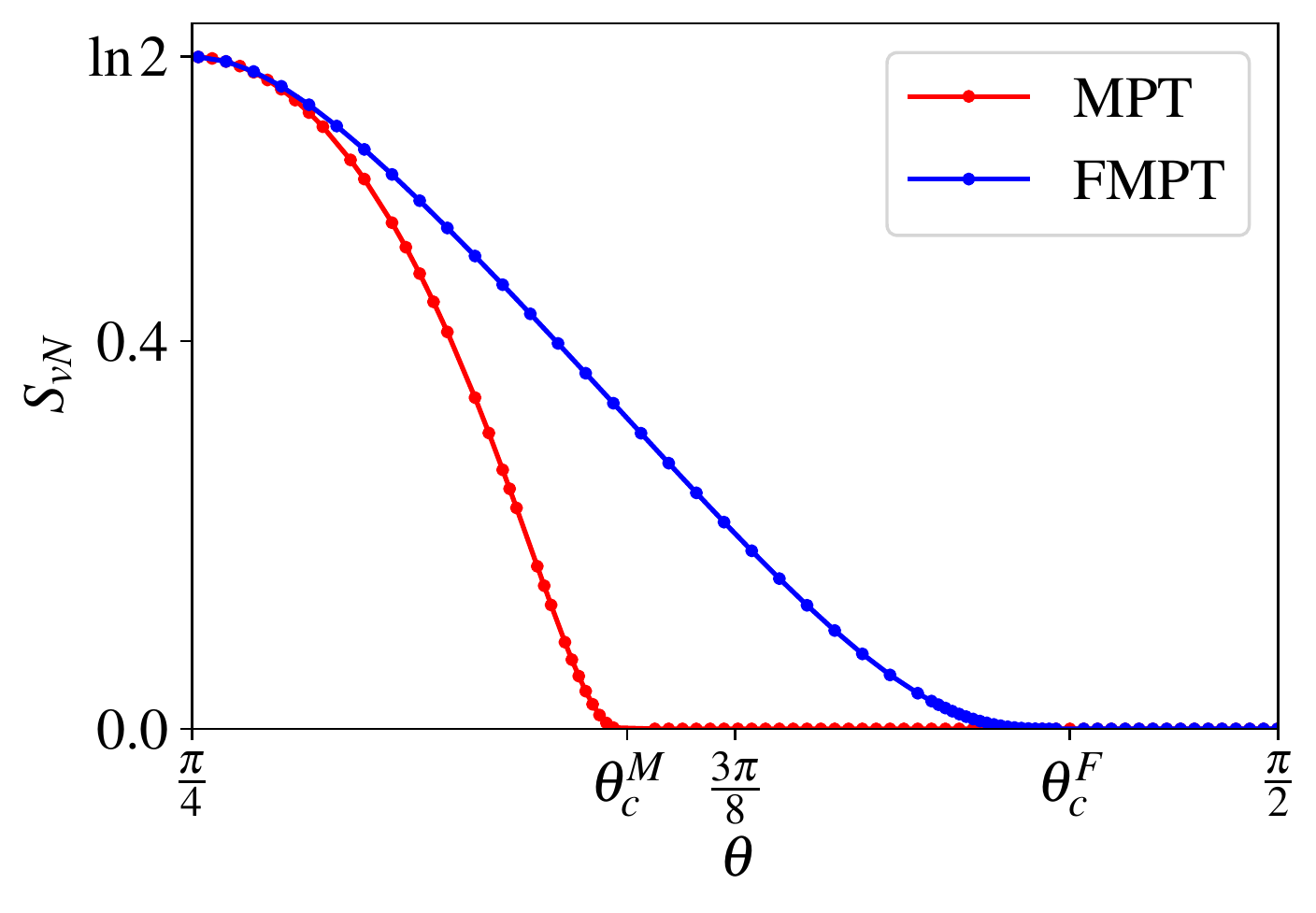}
    \caption{The purity of the state of the final spin in a collapse process, as characterized by the von Neumann entropy $S_\text{vN}$ of the corresponding density matrix. Here we plot the value of $S_\text{vN}$ in the limit of long time (large initial system size) for both the real measurement (MPT) and forced measurement (FMPT) cases. The horizontal axis shows the parameter $\theta$ that defines the strength of the weak measurement within each node of the tree (see Sec.~\ref{sec:collapse}); larger $\theta$ corresponds to stronger measurement. The calculated critical values for the MPT and FMPT cases are shown as $\theta^M_c$ and $\theta^F_c$, respectively. Symbols correspond to results from numerical calculations, and error bars are smaller than the symbol size. } 
    \label{fig:late_time_entropy}
\end{figure}

We find that the MPT and FMPT have qualitatively similar critical behavior, with an exponential vanishing of the order parameter as the transition is approached from the weak-monitoring side.  
However they have different values of the critical measurement strength, with the MPT exhibiting a smaller entangling phase than the FMPT. This difference can be seen in Fig.~\ref{fig:late_time_entropy}, which shows the von Neumann entropy $S_\text{vN}$ of the final state
(i.e. the final, unmeasured spin at the apex of the tree in Fig.~\ref{fig:collapse_expansion})
for a collapse process in the limit of long time (large initial system size). In the strong monitoring phase, $S_\text{vN} \rightarrow 0$ in the limit of long time, while in the weak monitoring phase $S_\text{vN}$ remains finite. For both types of dynamics, $S_\text{vN}$ is plotted against a parameter $\theta$ (defined below) that characterizes the strength of weak measurements within each node of the tree; $\theta \rightarrow \pi/4$ corresponds to the limit of no measurement while $\theta \rightarrow \pi/2$ is the limit of strong, projective measurements.

Interestingly, the MPT and FMPT reside in distinct universality classes, with the MPT (but not the FMPT) located at the boundary separating two different regimes for the phase transition. These two regimes can be thought of as ``strong'' versus ``weak'' randomness. 
This result suggests that there may be slightly different protocols
for which the MPT and FMPT exhibit more radical differences in their critical scaling, for example power-law versus exponential scaling of the order parameter.

The structure of this paper is as follows. 
In Sec.~\ref{sec:model} we define the dynamical tree models.
In Sec.~\ref{sec:force} we study the forced measurement case using both simulations and theoretical analysis. In Section~\ref{sec:real} we turn to  the case of real measurements, and contrast it with the forced measurement case. 
In Sec.~\ref{sec:expansion} we discuss the expansion process, and suggest a protocol for realizing the measurement phase transition on a tree experimentally. 
We conclude in  Sec.~\ref{sec:concl} with a summary and a discussion of  potential future work.

\section{Dynamical tree models}
\label{sec:model}

The quantum dynamics we consider are of two kinds.
In the \textit{collapse} process we start with a large number of spins (qubits).
In each timestep the number of ``active'' spins is halved, because, after the spins interact in pairs, half of them are projectively measured and discarded (Fig.~\ref{fig:collapse_expansion}). 
This process gives a tree structure in spacetime, with only a single spin remaining at the top of the tree, 
i.e.\ at the final time.
The question we ask is whether the state of this final spin  preserves quantum information from the initial state \cite{Gullans_Dynamical_2020}. 
This question may be formalized by initializing the spins in the lowest layer of the tree in the maximally mixed state and asking whether the final spin is in a pure state, or whether it is in a mixed state with a nonzero entropy (which can be at most $\ln 2$, since the final spin is a 2-state system). This entropy is a measure of the amount of surviving information, and is also an order parameter for the phase transition.

The \textit{expansion} process is, loosely speaking, the inverse process.  
Initially the ``system'' consists of a single spin. 
Further spins (initialized in a definite state,  $\ket{\uparrow}$) are recruited at each timestep so that at time $t$ the system consists of $2^t$ spins. Again we ask whether  a maximally mixed initial state is purified or not. 

An equivalent way to formulate this question about purification \cite{Gullans_Scalable_2020} in the expansion process is to start with a state in which the initial spin is maximally entangled with  a ``reference'' spin. In this case entropy of interest is also equal to the entanglement, at the final time, between the reference spin and the collection of $2^t$ system spins (which is again no larger than $\ln 2$).

The detailed dynamics are specified below and include unitary interactions and weak measurements.
Equivalent models could also be formulated using only projective measurements, at the cost of a more complex node interaction involving ancilla spins (a trivial way to see this equivalence is by using the physical interpretation of a weak measurement in terms of an ancilla\footnote{A weak measurement can be achieved by allowing the spin of interest to interact for a finite time with an ancilla degree of freedom, and then projectively measuring the ancilla \cite{clerk2010introduction}.}).

From a formal point of view, the basic  objects describing the evolution are tree tensor networks, built up (as shown in Fig.~\ref{fig:tree_model} for the collapse process) from: 
unitary gates $U$;  Krauss operators $K$ (representing weak measurements, see below); and kets/bras associated with the spins that are recruited/discarded in each timestep.
The set of allowed tree tensor networks is identical for both the collapse and expansion processes, just with the time coordinate reversed. 
In both cases the order parameter is a simple property of the singular value decomposition of the tree (for a decomposition separating the ``trunk'' from the ``leaves'').
However, the \textit{probability distribution} on this set of tree tensor networks depends crucially on the physical interpretation of the dynamics, because applying Born's rule (the usual quantum mechanical probability for measurement outcomes) leads to correlations between the constituent elements of the tensor network. 
We now describe the dynamics more carefully.

\subsection{The collapse process: basic ``node''}
\label{sec:collapse}

\begin{figure}[tb]
    \centering
    \includegraphics[width = 0.85 \columnwidth]{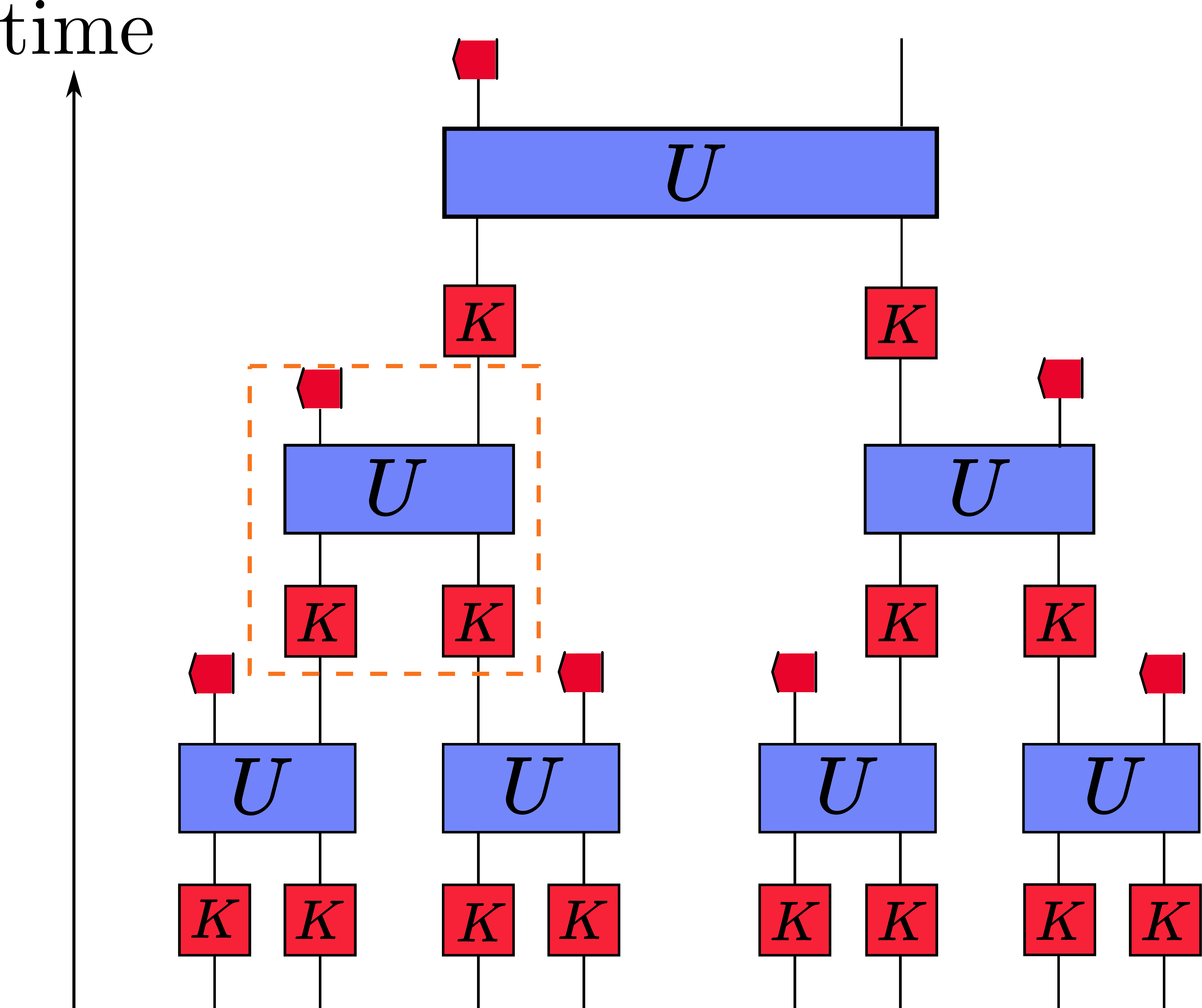}
    \caption{
    An illustration of the tree model studied in this paper (for the collapse process), drawn here with $k=3$ generations. Each line represents the world line of one of the $2^k$ spins in the system. The red $K$ blocks (defined by Eq.~\ref{eq:Kraus}) represent single-spin weak measurements. The blue $U$ blocks represent independent Haar-random, two-spin unitary operators. The red bras are associated with projective (strong) single-spin measurements. A single node of the tree (a circle in Fig.~\ref{fig:collapse_expansion}), comprising all three operations, is indicated by the orange dashed box.}
    \label{fig:tree_model}
\end{figure}

The basic ingredient of the collapse process (i.e. a node of the tree) 
is a random operation that takes a state of two spins as the input and gives a state of a single spin as output.
The node consists of the following sequence of three steps: 

(i) a weak measurement is applied independently to each of two spins along the lower bonds,

(ii) a Haar-random, two-spin unitary operator $U$ is applied to the two spins, and

(iii) one of the two spins undergoes a projective measurement.

The Haar-randomness of the two-site unitaries means that it does not matter whether we choose to make all the measurements in the $S_z$ basis, or whether we choose a new random basis for each measurement.
These two measurement protocols are related by a ``gauge transformation'' which involves rotating the measurement bases and making compensating rotations of the 2-site unitaries (see Appendix~\ref{app:gaugetransf} for more details).\footnote{Here we assume that the initial state is invariant under single-site unitary rotations, as for the maximally mixed case.} 
For notational convenience later on, 
we will take each weak measurement to be in an independent random basis,
but the strong measurements to be in the $S_z$ basis.

The unmeasured spin that remains after (i)--(iii) is then used as an input for another node of the tree, unless it is the final spin at the apex of the tree. The resulting tree is depicted in Fig.~\ref{fig:tree_model}.
The strength of the weak measurement in step (i) is the major parameter of the model.

These ingredients may be formalised as follows.
Each weak measurement is described using a Kraus operator 
\begin{align}
    K_{\sigma} = \cos \theta \mathds{1} + (\sin\theta-\cos\theta) \,
    u \ket{\sigma}\bra{\sigma} u^\dag, 
    \label{eq:Kraus}
\end{align} 
where ${\sigma = \uparrow}$ or $\downarrow$ labels the measurement outcome,   $\ket{\sigma}$ denotes an $S_z$ eigenstate,
and $u$ is a Haar-random unitary rotation that randomizes the measurement basis.
(An independent random basis is chosen for \textit{each} weak measurement, but 
to avoid clutter in our notation we will leave the dependence of $K_\sigma$ on $u$ implicit.)
If a given spin has the initial $2\times 2$ density matrix $\rho$, then after a weak measurement with outcome $\sigma$ the spin has the normalised density matrix
\be
\frac{K_{\sigma}\rho K_{\sigma}^{\dagger}}{\tr K_{\sigma}\rho K_{\sigma}^{\dagger}}.
\ee
If the weak measurement is a ``true'' measurement, the probability of the outcome $\sigma$ is equal to 
\be\label{eq:bornruleweak}
p_\sigma = \tr K_{\sigma}\rho K_{\sigma}^{\dagger}.
\ee
For a forced measurement, we fix the outcome to $\sigma = \uparrow$, without loss of generality. See the following subsection for a comment on the physical interpretation of forced measurements.

The strength of the weak measurement is defined by the parameter $\theta$,
\be
\theta\in \left[ {\pi}/{4}, {\pi}/{2} \right].
\ee
When $\theta = \pi/4$ the Kraus operator is  proportional to the identity and no measurement occurs,  while when ${\theta = \pi/2}$ the Kraus operator is a projection operator, giving a conventional projective measurement. For both the MPT and the FMPT we will find that the critical value of $\theta$ lies in the interior of this interval, with ${\theta=\pi/4}$ being in the entangling phase and $\theta=\pi/2$ in the disentangling phase.

Let the initial density matrices of the (uncorrelated) spins involved in the node be $\rho_1$ and $\rho_2$. Then the action of the weak measurements is
\begin{align}
\rho_1\otimes \rho_2 & \longrightarrow \rho' \equiv 
\frac{K^{(1)}_{\sigma_1}\rho_1 K_{\sigma_1}^{(1)\dagger}}{\tr K^{(1)}_{\sigma_1}\rho_1 K_{\sigma_1}^{(1)\dagger}}\otimes \frac{K^{(2)}_{\sigma_2}\rho_2 K_{\sigma_2}^{(2)\dagger}}{\tr K^{(2)}_{\sigma_2}\rho_2 K_{\sigma_2}^{(2)\dagger}},
\end{align} 
with the appropriate probabilities for the outcomes $\sigma_1$, $\sigma_2$.
(We write the superscripts $(1)$ and $(2)$ for the Kraus operators because $K^{(1)}_\uparrow$ --- for example --- differs from $K^{(2)}_\uparrow$ as a result of the independent choices of random measurement basis for each spin).
Next, in step (ii) we generate a $4 \times 4$ Haar-random unitary matrix $U$ and act on the combined state of the two spins so that the new $4\times 4$ density matrix is
\be\label{eq:rhopp}
\rho'' = U\rho' U^{\dagger}. 
\ee
Finally, in step (iii) we perform a projective measurement of one spin (taken here to be spin 2, without loss of generality)  in the $S^z$ basis.
For measurement outcome $\sigma_2'$
(which can again either be selected with Born's rule, or postselected to enforce $\sigma_2'= \uparrow$) 
 the corresponding projection operator is 
${P_{\sigma_2'} = \mathbf{1}\otimes \lf \ket{\sigma'_2}\bra{\sigma'_2} \ri}$,
and the final $2\times 2$ density matrix for the first spin after all three steps~is

\be
 \rho''' = \frac{\tr_{2} P_{\sigma'_2}\rho'' P_{\sigma'_2}}{\tr P_{\sigma'_2}\rho''P_{\sigma'_2} }
 \label{rho_f}. 
\ee
Here $\tr_2$ denotes the partial trace over the second spin, while $\tr$ denotes the trace over both spins.

For real measurements, the joint probability for  ${p(\sigma_1, \sigma_2, \sigma_2')}$,
given the   initial state and the random unitaries,
can be written as the product of the probabilities for the earlier measurements and the conditional probability for the later one:
\be
p(\sigma_1,\sigma_2,\sigma'_2) =p^{(2')}(\sigma'_2|\sigma_1,\sigma_2)\, p^{(1)}(\sigma_1)\,p^{(2)}(\sigma_2).
\ee
Using Eq.~\ref{eq:bornruleweak}, and  ${p^{(2')}(\sigma'_2|\sigma_1,\sigma_2)= \tr P_{\sigma_2'} \rho''  P_{\sigma_2'}}$,
gives
\begin{align} \notag
    p(\sigma_1,\sigma_2,& \sigma'_2)=  \\ 
    \tr  & \Big[  P_{\sigma'_2} U \lf {K^{(1)}_{\sigma_1}\rho_1K_{\sigma_1}^{(1)\dagger}\otimes K^{(2)}_{\sigma_2}\rho_2K_{\sigma_2}^{(2)\dagger}} \ri U^{\dagger} P_{\sigma_2'} \Big]. \notag
\end{align}
Anticipating the notation of the next section, the conditional probability can be written more conveniently in terms of a node tensor $t$,
\be
p(\sigma_1, \sigma_2, \sigma_2' )  =
\tr \, t(\sigma_1, \sigma_2, \sigma_2' ) (\rho_1 \otimes \rho_2) t(\sigma_1, \sigma_2, \sigma_2' )^\dag. \label{eq:real_outcomep}
\ee
Similarly, the output density matrix is 
\be
\rho''' = \f{
t(\sigma_1, \sigma_2, \sigma_2' ) (\rho_1 \otimes \rho_2) t(\sigma_1, \sigma_2, \sigma_2' )^\dag
}{\tr \, t(\sigma_1, \sigma_2, \sigma_2' ) (\rho_1 \otimes \rho_2) t(\sigma_1, \sigma_2, \sigma_2' )^\dag}.
\label{eq:final_density_matrix}
\ee

\subsection{Collapse process as a tensor network}\label{sec:collapseTN}
 
The full collapse process can be described with the tensor network $T$ which is shown in Fig.~\ref{fig:tree_model}. 
This tensor network can be viewed as a $2\times M$ rectangular matrix, where the number of rows, 2, is the Hilbert space dimension of the final remaining spin (the topmost bond of $T$) and the number of columns,   ${M=2^{2^k}}$, is the Hilbert space dimension of the $2^k$ initial spins (the lowermost bonds). $T$ depends on the various random variables as follows.

First, there are the various random unitaries. We denote the complete set of all the unitaries appearing in $T$ by $\mathbb{U}$. 
For each node $r$ there is a two-site random unitary $U_r$, 
together with two single-site random unitaries,
$u^{(1)}_r$ and $u^{(2)}_r$, 
that appear inside the Kraus operators and set the weak measurement bases.
Second, $T$ depends on the lists,  
denoted ${\bf m}_W$ and ${\bf m}_S$, of measurement outcomes for the weak and strong measurements respectively. 
We will sometimes make the dependence on the measurement outcomes explicit,  $T=T({\bf m}_W,{\bf m}_S)$,
but we will leave the dependence of $T$ on the various unitaries implicit.
We will do the same for the local node tensors $t$ described below.

The tensor network $T$ is built by contracting together three-index tensors $(t_r)^a_{bc}$, where $r$ labels a trivalent node of the tensor network
(from now on we will suppress this subscript)
and $a$, $b$, $c$ are the bond indices. 
The node tensor is constructed  as follows.
First, we define  
\be
\widetilde t(\sigma_1, \sigma_2) =  U \, (K^{(1)}_{\sigma_1}\otimes K^{(2)}_{\sigma_2}), \label{tilde-t-tensor}
\ee 
which is an operator acting on two spins. We write its components as $\widetilde t^{ab}_{cd}$, where the positions of the indices reflects their geometrical position in the tensor network: 
the indices $(a,b)$ are for the upper bonds of the tensor, and 
$(c,d)$ are for the lower bonds. 
Viewed as a matrix, the multi-index $(a,b)$ makes up the row index of $\widetilde t$ and $(c,d)$ makes up the column index.

The strong measurement outcome fixes the value of the index $b$, so the desired three-index tensor, conditioned on all three measurement outcomes, is:
\be\label{eq:t3index}
[t(\sigma_1, \sigma_2, \sigma_2')]^a_{cd}
=
\widetilde t(\sigma_1, \sigma_2)^{a, \sigma_2'}_{cd}.
\ee

If we take the $2^k$ spins that comprise the initial state to be in a maximally mixed state, so that the initial density matrix of the system is proportional to the identity  matrix, $\rho = \mathds{1} / M$, then
the final state of the last spin can be written  as (again we make the dependence on the measurement outcomes explicit):
\be
\rho_f({\bf m}_W,{\bf m}_S)  =\f{ T({\bf m}_W,{\bf m}_S) T({\bf m}_W,{\bf m}_S)^\dag }{\tr \, T({\bf m}_W,{\bf m}_S) T({\bf m}_W,{\bf m}_S)^\dag}.
\ee
The matrix multiplication in $TT^\dag$ contracts all the bottom bonds of the tensor network $T$ (Fig.~\ref{fig:tree_model}) with the corresponding bonds of $T^\dag$,
leaving a $2\times 2$ matrix.

We now consider the probability distribution for the tensor network $T$ for protocols involving either forced or true measurements.

In all cases, the unitaries in the set  $\mathbb{U}$ are drawn independently from the Haar distribution. 
In the \textit{forced} measurement case, this set of unitaries is sufficient to fix $T$ entirely. 
Using the unitaries we have drawn, we can form the local node tensors 
$t=t(\uparrow, \uparrow, \uparrow)$.

Note that in this forced measurement case the local $t$ tensors for  the nodes are independently random, 
i.e.\ they are uncorrelated between nodes.\footnote{ 
Physically, we could sample the forced measurement trees using the following procedure. We first choose $\mathbb{U}$, then attempt to run the dynamics. If the dynamics produces a ``wrong'' measurement outcome, i.e. $\downarrow$, we discard that run and re-run the dynamics (in fact it is sufficient to re-run the relevant subtree)  \textit{using the same $\mathbb{U}$}. We repeat this process until we get an all-$\uparrow$ measurement record for the given  $\mathbb{U}$. Then we resample $\mathbb{U}$ and repeat. Note that this protocol does not bias the distribution of $\mathbb{U}$: all unitaries are Haar-distributed.}
In the true measurement case the local measurement outcomes, and therefore the local tensors $t=t(\sigma_1, \sigma_2, \sigma_2')$, are correlated.
 
For \textit{true} measurements,
the tree
$T({\bf m}_W,{\bf m}_S)$
is determined  by both $\mathbb{U}$ and the measurement trajectory. 
The probability distribution for $T({\bf m}_W,{\bf m}_S)$ is the product of the Haar distribution for $\mathbb{U}$ and
the conditional probability (given $\mathbb{U}$) of obtaining the trajectory  $({\bf m}_W,{\bf m}_S)$.
The probability of such a trajectory may be written as
\be\label{eq:ptrueintermsofT}
p({\bf m}_W,{\bf m}_S) =  \f{1}{M}  \tr \, T({\bf m}_W,{\bf m}_S) T({\bf m}_W,{\bf m}_S)^\dag.
\ee
In practise, rather than using this expression for the full trajectory, it will be more convenient to think of the process as a Markov process, in 
which at each time step the density matrices are acted on in the manner described in the previous section.

We have distinguished the case where all measurements are true from the case where all measuremements are forced.  
We will also mention a mixed case, where the weak measurements are true measurements, but the strong measurements are postselected to be $\uparrow$.
In this case, the probability for a given tree is determined by conditioning
the probability Eq.~\ref{eq:ptrueintermsofT}  
on having ${{\bf m}_S={\bf 0}}$, 
where  ${\bf 0}$ is the list  ${(\uparrow, \uparrow, \ldots, \uparrow)}$ of length ${2^k-1}$. 
This process gives\footnote{
In more detail: ${p_\text{mixed}({\bf m}_W)}$ is the conditional probability
${\operatorname{Prob}({\bf m}_W | {\bf m}_S={\bf 0})}$.
In terms of $p({\bf m}_W, {\bf m}_S)$ in Eq.~\ref{eq:ptrueintermsofT}, this conditional probability is 
${p_\text{mixed}({\bf m}_W) = 
{p({\bf m}_W, {\bf 0})}/ ( \sum_{{\bf m}_W'} p({\bf m}_W', {\bf 0}) )}$.
Using (\ref{eq:ptrueintermsofT}) and repeatedly using the relation ${\sum_\sigma K_\sigma K_\sigma^\dag=\mathds{1}}$ for the Kraus operators, the denominator simplifies to $2^{1-M}$.}
\be\label{eq:pmixedexpr}
p_\text{mixed} ({\bf m}_W) = \f{1}{2} 
\tr \, T({\bf m}_W, {\bf 0} ) T({\bf m}_W,  {\bf 0} )^\dag
\ee
(where the nontrivial probability distribution is only for ${\bf m}_W$, since ${\bf m}_S$ is fixed). As we will discuss, the distribution of weak measurement outcomes on the right-hand side of 
Eq.~\ref{eq:pmixedexpr} is also precisely the one relevant to the expansion process with true (weak) measurements.

\subsection{Characterizing the transition}

We adopt the perspective mentioned in the introduction, and introduced in Ref.~\cite{Gullans_Dynamical_2020}, of whether a maximally-mixed initial state is purified by the dynamics. 
This is equivalent to studying the singular value decomposition of the tree tensor network/rectangular matrix~$T$.

As above, the  $2^k$ spins in the collapse process are initially in a maximally mixed state.
We ask whether the density matrix $\rho_f$ of the unmeasured spin at the apex of the tree is in a pure state or a mixed state.
In its eigenbasis, $\rho_f$ may be written
\begin{align}
    \rho_{f} & = \begin{pmatrix}
    1-Z & 0\\
    0 & Z
    \end{pmatrix},
    &
    0\leq Z \leq 1/2.
      \label{eq:rho}
\end{align}
The value of  $Z$, the smaller of the two eigenvalues of $\rho_f$, determines how mixed the state is.
$Z = 0$ corresponds to a pure state, 
while $Z = 1/2$ describes a maximally mixed state. 
In the near-critical regime, where $Z$ is small, the R\'enyi entropies are approximately ${S_\mathrm{vN} \equiv  S_1 = \simeq Z \ln Z^{-1}}$, and $S_n \simeq n(n-1)^{-1} Z$ for $n>1$.

We use the notation $Z_k$ to denote the smaller eigenvalue of a tree of $k$ generations.
In the limit of large $k$, $Z_k$ yields an order parameter for the MPT or FMPT: the entangling phase 
is defined by the fact that the typical (or the average) value of $Z_k$ is nonzero in the limit ${k\rightarrow \infty}$.
As discussed above, $Z$ is closely related to the entropies, so that having finite $Z$ in the limit ${k \rightarrow \infty}$ is equivalent to having a nonzero entropy in the limit of a large tree.
We have already shown numerical data for the average entropy ${\langle S_\mathrm{vN}\rangle}$ in Fig.~\ref{fig:late_time_entropy}, 
illustrating the critical vanishing at the relevant critical value of $\theta$.

The evolution of the distribution of $Z_k$ as a function of $k$ can be described recursively, as detailed in the following sections. 
For an analytical treatment it will be convenient to focus on the \textit{typical} value of $Z_k$ \cite{Derrida_Polymers_1988},
defined by\footnote{In Ref.~\cite{Nahum_Measurement_2021}, care was needed to take the average in Eq.~\ref{eq:ztyp} over only nonzero values of $Z_k$. In the model we study here, however, the value of $Z_k$ is always nonzero for all finite $k$ unless $\theta = \pi/2$.}
\be 
\ln \ztyp_k = \langle\ln Z_k \rangle,
\label{eq:ztyp}
\ee 
instead of the mean value $\langle Z_k \rangle$.
This distinction is important because $Z_k$ may have a broad statistical distribution for trees of size ${k \gg 1}$, so that ${\langle Z_k \rangle}$ is dominated by rare samples and is much larger than the typical value. 
(Nevertheless,  ${\lim_{k\rightarrow\infty}\ztyp_k}$ and ${\lim_{k\rightarrow\infty} \langle Z_k\rangle}$ are both nonzero precisely in the entangling phase.)

To recap, the phenomenology of the transition is that in the entangling phase the typical value $\ztyp_k$ tends to a  positive constant in the limit $k \rightarrow \infty$, indicating that a nonzero entropy is retained in the final state, while in the disentangling phase, $\ztyp_k\rightarrow 0$ as $k\rightarrow\infty$.

\subsection{Expansion process}
\label{sec:defineexpansionprocess}

The expansion process at time ${t=k}$ (Fig.~\ref{fig:collapse_expansion}, Right) is characterized by a tensor network that relates the state of a single initial spin to the state of all $2^k$ spins at the final time. (This final state includes the $2^k - 1$ spins that were  recruited during the $k$ timesteps.)

The relevant tensor networks are related to the  ones in the collapse process, defined in Sec.~\ref{sec:collapseTN}, by reversing the direction of the arrow of time.

For the case of forced measurements, where all measurements are postselected to be  $\uparrow$,
the expansion process is precisely equivalent to the collapse process except for this time reversal.
A node in the expansion process which ``recruits'' a spin in the state $\ket{\uparrow}$ can be viewed as the time reversal of a node in the collapse process that discards a spin after a forced measurement in the state $\ket{\uparrow}$.
More precisely, the relevant ensemble of tensor networks is the exactly same between the two forced measurement processes 
(with the probability distribution given simply by the Haar distributions for the unitaries). Since the FMPT is a property of the ensemble of tensor networks, it follows that the expansion process has all the same critical properties as the collapse process in the FMPT case.

In the expansion process with true measurements, the weak measurement outcomes ${\bf m}_W$ are sampled with Born's rule (there are no strong measurements in the expansion process). 
This process is \textit{not} equivalent to the collapse process with only true measurements. Instead, writing the relevant probability ${p_\text{expansion}({\bf m}_W)}$ in terms of $T$
shows that ${p_\text{expansion}({\bf m}_W)}$  is equal to the probability  
${p_\text{mixed}({\bf m}_W)}$
 of the ``mixed'' collapse process in Eq.~\ref{eq:pmixedexpr}, where the weak measurements are true but the strong ones are forced.

In our discussion of the MPT in Sec.~\ref{sec:real} we focus on the collapse process. We comment in Sec.~\ref{sec:expansion} on an efficient experimental protocol for detecting the transition in an expansion process.

\section{Forced measurements}
\label{sec:force}

Before considering real measurements, in this section we examine the case of forced measurements, as these can be understood by a straightforward application of the approach in Ref.~\onlinecite{Nahum_Measurement_2021}. We  briefly recapitulate this approach,  then give the resulting critical properties for the FMPT. As mentioned above, in the forced measurement case all measurement outcomes are fixed in advance to give the $\uparrow$ state in the corresponding measurement basis.

Below we discuss the collapse process FMPT.
However the results below also apply for the expansion process FMPT, for the reason described in 
Sec.~\ref{sec:defineexpansionprocess}: the same ensemble of tensor networks describes both problems.

\subsection{Linearized Recursion Relation}
\label{sec:force_linear}

In order to identify a recursion relation for the probability distribution of the smaller eigenvalue $Z$ (see Eq.~\ref{eq:rho}), we imagine the process of constructing a tree with $k+1$ generations from two smaller subtrees, each with $k$ generations.  
We denote the probability distribution of the smaller eigenvalue for a tree of  $k$ generations by $\mu_k(Z)$.

Suppose that the two subtrees have final states that are described by density matrices $\rho_{k, 1}$ and $\rho_{k,2}$, respectively, with corresponding smaller eigenvalues $Z_{k,1}$ and $Z_{k,2}$. 
Since the two subtrees are statistically independent, each of these eigenvalues is drawn independently from the distribution $\mu_k$. 
Combining these trees into a larger tree with $k+1$ generations via the  ``node'' operation described in Sec.~\ref{sec:collapse}, with the appropriate probabilities for the unitaries and measurements,
gives a new final density matrix $\rho_{k+1}$ whose smaller eigenvalue
 $Z_{k+1}$ is distributed according to $\mu_{k+1}$.

In considering this process of combining subtrees,
it is in fact sufficient to take
$\rho_{k, 1}$ and $\rho_{k,2}$ to be diagonal and of the form in Eq.~\ref{eq:rho}. 
This simplification arises because the unitary transformations required to achieve this diagonal form can be absorbed into the Haar-random unitaries involved in the node tensor,
without changing the probability distribution of these operations.
This is a crucial simplification from Haar randomness:
it means we can deal with a recursion relation only for the eigen\textit{value} $Z$, without having to keep track of how the associated eigen\textit{vectors} of the density matrix evolve.

In a given instance, 
the new eigenvalue $Z_{k+1}$  depends on the values of $Z_{k, 1}$ and $Z_{k, 2}$, as well as on the variables in the node ($U$, and the random basis rotations in $K^{(1)}$ and $K^{(2)}$). 
For example, if $Z_{k, 1} = Z_{k, 2} = 0$, then we necessarily have $Z_{k+1} = 0$, since the action of the node always takes two independent pure states into a single pure state. 

In order to understand the critical properties near the transition, it is sufficient to consider the case where $Z_k$ is very small compared to unity (either vanishing at large $k$ in the disentangling phase, or tending to a very small constant just inside the entangling phase). 
Much of the key information is contained in the leading order (linear) expansion of $Z_{k+1}$ in terms of  $Z_{k, 1}$ and $Z_{k,2}$:
\be\label{eq:Zlinear}
Z_{k+1} = A_1 Z_{k,1} + A_2 Z_{k,2} + O(Z_k^2).
\ee
For a given node, the coefficients $A_1$ and $A_2$ 
can be expressed in terms of the three-index tensor $t$ defined in Sec.~\ref{sec:collapseTN},
which makes up a local node of the tree tensor network. 
For the true measurement problem, this tensor depends on the measurement outcomes at the node, 
\be
t = t(\sigma_1,\sigma_2, \sigma_2'), 
\ee
while for the forced measurement problem these are fixed, and 
\be
t = t(\uparrow,\uparrow, \uparrow).
\ee
The tensor $t$ is therefore defined by the Haar-random 2-site interaction unitary and the two Haar-random 1-site unitaries appearing in the Kraus operators  (cf. Eqs.~\ref{tilde-t-tensor},~\ref{eq:t3index}):
\be\label{eq:repeatt}
t^{a}_{cd}= \big[ U \, (K^{(1)}_{\uparrow}\otimes K^{(2)}_{\uparrow}) \big]^{a \uparrow}_{cd}.
\ee   
In terms of $t$, Eq.~\ref{rho_f} becomes
\be\label{eq:nonlinearrecursionrho}
\rho_f = \f{t
( \rho_{k,1}\otimes\rho_{k,2})t^{\dagger}
}{\tr\, t
( \rho_{k,1}\otimes\rho_{k,2})t^{\dagger}},
\ee 
where  $t^a_{cd}$ is treated as a $2\times 4$ rectangular matrix whose row index is $a$ and whose column index is $(c,d)$.

From this expression for $\rho_f$ we can write the expression for $\tr \rho_f^2$, and the latter can be expanded in $Z_{k+1}$ as 
${\tr \rho_f^2\simeq 1-2Z_{k+1}}$.
Explicitly evaluating the left-hand side gives the random coefficients in (\ref{eq:Zlinear}) as:
\begin{align}
    &A_1= \frac{\left|t_{11}^{1}t_{21}^{2}-t_{21}^{1}t_{11}^{2}\right|^2}{\left(|t_{11}^{1}|^2+|t_{11}^{2}|^2\right)^2}, \label{linear_coeff1}\\
    &A_2=\frac{\left|t_{11}^{1}t_{12}^{2}-t_{12}^{1}t_{11}^{2}\right|^2}{\left(|t_{11}^{1}|^2+|t_{11}^{2}|^2\right)^2}. \label{linear_coeff2}
\end{align}

\begin{figure}[t]
    \centering
    \includegraphics[width = 0.9\columnwidth]{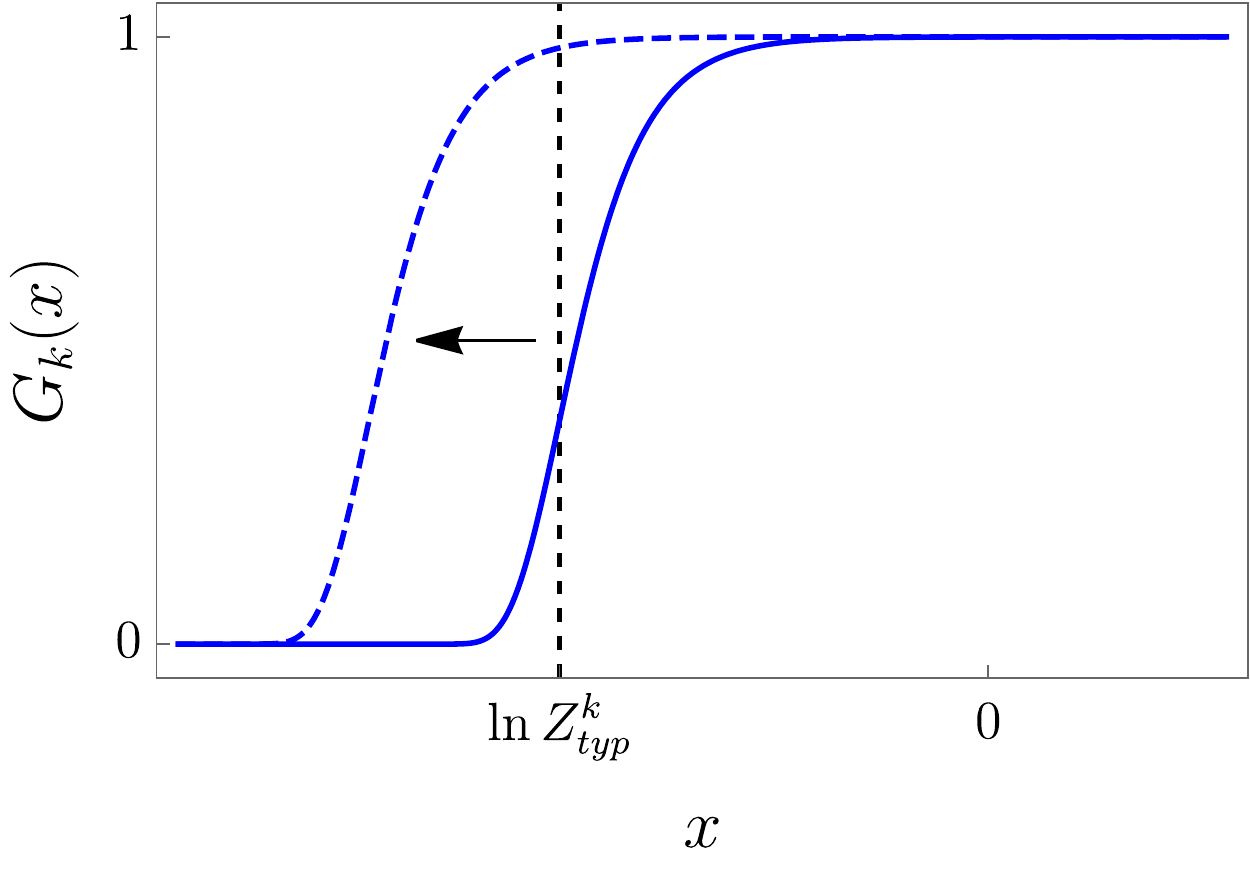}
    \caption{Schematic illustration of the generating function $G_k(x)$ (solid blue curve). The value of $\ln \ztyp_k$ corresponds to the position of the wave front.  In the disentangling phase, this front moves toward negative $x$ with increasing time (increased $k$).}
    \label{fig:g_function}
\end{figure}

In order to understand the evolution of $Z_k$ with increasing $k$, we define the generating function \cite{Derrida_Polymers_1988}
\be
G_k(x) = \langle \exp(-e^{-x}Z_k) \rangle. 
\label{eq:Gdef}
\ee
Here, $\langle ... \rangle$ indicates the average over $Z_k$ with distribution $\mu_k$. 
A heuristic way to think of $G_k(x)$ is as a smeared version of the cumulative probability distribution of $\ln Z_k$:
For ${x \gg \ln \ztyp}$, $G_k(x)$ has a plateau at unity, and for 
 ${x \ll \ln \ztyp}$ it has a plateau at zero,
as illustrated in Fig.~\ref{fig:g_function}.

As in Ref.~\cite{Nahum_Measurement_2021}, we first study the linearized recursion relation (Eq.~\ref{eq:Zlinear}), and we then take into account the nonlinear terms.
We are interested in the distribution of $Z_k$ for large $k$, i.e.\ in the nature of the probability distribution $\mu_k$ after many recursive steps.

The linearized recursion for $Z$ (Eq.~\ref{eq:Zlinear}) implies a  recursion relation for $G_k(x)$ that can be expressed through the coefficients $A_1$ and $A_2$ as:
\be
G_{k+1}(x) = \langle G_k(x-\ln A_1)G_k(x-\ln A_2)\rangle \label{generator_recursion},
\ee
where the remaining average is over $A_1$ and $A_2$.
Note that the probability distribution of $A_{1,2}$ depends on the measurement strength  parameter $\theta$ of our model. 

It is convenient to view $k$ as the time and $x$ as a kind of position for a travelling wave \cite{Derrida_Polymers_1988},
as suggested by Fig.~\ref{fig:g_function}.
Then Eq.~\ref{eq:Gdef} describes a wave front that is located at ${x = \ln \ztyp_k}$ and which propagates in ``space'' with a velocity $v_\theta$ that may be either positive or negative. 
The critical point $\theta_c$ corresponds precisely to the value for which $v_\theta$ vanishes \cite{Nahum_Measurement_2021}.

In the disentangling phase, the wave front propagates  to the left ($v_\theta<0$)  at large times $k$, meaning that the  typical value of $Z_k$ decays exponentially, ${\ztyp\sim e^{-|v_\theta|k}}$.
In the entangling phase, on the other hand,
the linearized treatment gives a traveling wave that propagates to the right ($v_\theta>0$).
Since a rightmoving wave would correspond to $\ztyp$ growing indefinitely, the nonlinear terms in  Eq.~\ref{eq:Zlinear} become important in the entangling phase: they halt the front at  finite value of $\ln \ztyp_k$ as discussed below in Sec.~\ref{sec:force_scale}.
However, our first task is to understand the linearized problem, which is captured by Eq.~\ref{generator_recursion}: from this linearized recursion we can identify the location of the critical point.

The travelling wave ansatz is 
\be
G_k(x) = G^{(\lambda)}\big(x-v_{\theta}(\lambda)k\big). 
\ee
Here $\lambda\geq 0$ parameterizes a family of possible travelling wave solutions: 
it will be necessary to choose the correct solution that is appropriate to the initial conditions. 
The value of $\lambda$ is defined through the exponential decay of  $G^{(\lambda)}$ at large argument (see Ref.~\onlinecite{Derrida_Polymers_1988} and references therein):
\be
G^{(\lambda)}(u) \simeq 1- e^{-\lambda u}. 
\label{eq:lambda-def}
\ee
Using this asymptotic expression and the recursion relation for $G_k(x)$ (Eq.~\ref{generator_recursion}) gives the following result for the wave front velocity:
\be
v_{\theta} (\lambda) = \frac{1}{\lambda}\ln \lf \left\langle A_1^{\lambda} \right\rangle
+ \left\langle A_2^{\lambda}\right\rangle \ri \label{eq:force_velocity}
\ee
where again the average is over $A_{1,2}$ as defined by Eqs.~\ref{linear_coeff1} and \ref{linear_coeff2}, or in other words over the random unitary operations that define the node tensor $t$.
Note that $t$, and therefore $A_{1,2}$, depend on $\theta$ implicitly through the Kraus operators in Eq.~\ref{eq:repeatt} [we will sometimes write ${A_{1,2}(\theta)}$ to emphasize this dependence].

The value of the parameter $\lambda$ is determined in a standard way as for Fisher-Kolmogorov-Petrovsky-Piskunov (FKPP) waves \cite{Fisher_The_1937}:
so long as the initial condition decays sufficiently fast at large $x$ (which we confirm below), the solution $\lambda$ that is selected is the one for which the velocity $v_\theta(\lambda)$ is minimal.
The ``dispersion relation'' $v_\theta(\lambda)$ is a convex function of $\lambda$, and  
for each fixed $\theta$ there is a value of $\lambda = \lambda^*$ that gives the minimal velocity: 
\begin{align}
\left.\frac{\partial v_{\theta}(\lambda)}{\partial \lambda}\right|_{\lambda^{\ast}} & =0,
&
v_{\theta} & \equiv v_{\theta}(\lambda^{\ast}).
\end{align}
If we plug in Eq.~\ref{eq:force_velocity}, we get
\be\label{eq:differentiatev}
-\frac{1}{\lambda^{\ast 2}}\ln (\langle A_1^{\lambda^{\ast}}+A_2^{\lambda^{\ast}}\rangle)+\frac{1}{\lambda^{\ast}}\frac{\langle A_1^{\lambda^{\ast}}\ln A_1+A_2^{\lambda^{\ast}}\ln A_2 \rangle}{\langle A_1^{\lambda^{\ast}}+A_2^{\lambda^{\ast}}\rangle} = 0.
\ee
For a given value of $\theta$, the above equation fixes $\lambda_*$ [and therefore ${v_\theta\equiv v_\theta(\lambda^\ast)}$].
At the critical point we have a second equation, namely
\be
v_{\theta_c}=0.\label{eq:viszero}
\ee
Notice that the first term on the left-hand side of Eq.~\ref{eq:differentiatev}
is equal ${v_\theta / \lambda}$, which vanishes at $\theta_c$.
Therefore the two equations which hold at the critical point, 
Eqs.~\ref{eq:differentiatev} and \ref{eq:viszero},
reduce to 
\begin{align}
\langle A_1^{\lambda^{\ast}}\ln A_1\rangle
+ \langle A_2^{\lambda^{\ast}}\ln A_2\rangle & =0. \label{eq:force_derivative} \\  \label{eq:secondcriteqn}
\langle A_1^{\lambda^\ast} \rangle
+ \langle A_2^{\lambda^\ast}\rangle  & = 1.
\end{align}
These two equations determine both unknowns, $\lambda^*$ and $\theta_c$, i.e. they determine the location of the critical point.

Surprisingly, Eqs.~(\ref{eq:force_derivative}) and (\ref{eq:secondcriteqn}) can be solved analytically.
Equation~\ref{eq:force_derivative} can be solved  for $\lambda^*$ using only an invariance property of the node tensor distribution (we consider this solution first), while for Eq.~(\ref{eq:secondcriteqn}) we must consider the particular structure of the node tensor.

Since both the weak measurement bases and the unitary operator $U$ are chosen Haar-randomly, the $t$-tensor is statistically invariant under  unitary rotations on any single index.
For instance, consider the mapping
\be\label{eq:tinvarianceproperty}
t^{a}_{cd} \to u_{cc'}\, t^{a}_{c'd};
\ee
the two tensors $t$ and $t' = ut$ have equal probability. 
Given this condition,\footnote{Here we also use the fact that $A_i>0$ with probability 1.} the coefficients $A_1$ and $A_2$ satisfy  
the identity \cite{Nahum_Measurement_2021} (irrespective of the value of $\theta$)
\be
\langle A_i^{1/2}\ln A_i\rangle = 0.
\label{eq:A12}
\ee
This may be shown by 
averaging over a single-leg unitary rotation like that in Eq.~\ref{eq:tinvarianceproperty}  (an  analogous argument is discussed in
App.~\ref{appendix:lambda}).
Given this identity,
Eq.~\ref{eq:force_derivative}  implies that ${\lambda^\ast = 1/2}$ at the critical point, and therefore that the critical point $\theta_c$ is the value of $\theta$ for which (Eq.~\ref{eq:secondcriteqn})
\be
 \langle A_1(\theta_c)^{1/2}
  \rangle
 + \langle 
 A_2(\theta_c)^{1/2}\rangle = 1,
\label{eq:AthetaFMPT}
\ee
where we have made the dependence of $A_i$ on $\theta$ explicit.
In the present model the two terms in 
Eq.~\ref{eq:AthetaFMPT} are equal by symmetry 
(the two bottom legs of $t$ are equivalent on average).

Equation~\ref{eq:AthetaFMPT}, together with Eqs.~\ref{linear_coeff1} and \ref{linear_coeff2}, defines the critical measurement strength $\theta_c$.
Equation \ref{eq:AthetaFMPT} is easy to solve numerically,  since it involves only a \textit{finite-dimensional} integral over the random parameters of a single node, but in fact it can also be solved analytically.

So far, the identities above relied only on the unitary invariance property
discussed around Eq.~\ref{eq:tinvarianceproperty}, and did not depend on the more detailed structure of the node tensor $t$.
For the present model, where $t$ is given by  Eq.~\ref{eq:t3index}, averages like $\langle A^\lambda\rangle$ can be written as explicit functions of $\theta$ by sequentially integrating out the various Haar-random objects appearing in the node.
This calculation is presented in Appendix~\ref{app:averages}
and gives
\begin{align} \notag
& \left\langle A_1^\lambda \right\rangle =
\left\langle A_2^\lambda \right\rangle =\\
&\,\,\, \f{3\times 4^{1-\lambda}
(\sin 2\theta)^{2\lambda}
(\cos2\theta)^{-1}
\lf  (\sin\theta)^{2-4\lambda} - (\cos\theta)^{2-4\lambda}   \ri
}{
(\lambda+1)(\lambda+2)(2\lambda-1)
(\lambda-2) (\lambda-3)
}.\label{eqn:analytical_a}
\end{align}
Taking the limit ${\lambda =1/2}$ in this formula, Eq.~\ref{eq:AthetaFMPT} becomes
\begin{align}
\f{\ln \gamma}{\gamma - 1/\gamma } & =
\f{75}{256}, &
&\text{for $\gamma=\tan\theta_c$},
\end{align}
which has the solution
\be
\label{eq:thetacFMPT}
\theta_c \simeq 1.42010054727632.
\ee
We also obtained $\theta_c$ by evaluating the averages in Eq.~\ref{eq:AthetaFMPT} directly, by numerically integrating over the Haar-random elements in the node, and we obtained the compatible result ${\theta_c = 1.4201\pm 0.0001}$.

Above we assumed that the travelling wave converged
to the solution with with $\lambda=\lambda^\ast$.
For completeness, let us briefly recall the justification for this assumption.

The  solution chosen at late times depends on which decays faster with $x$ at large $x$:
the initial condition ${1-G_0(x)}$,
or the minimal-velocity wavefront, ${1-G^{(\lambda^\ast)}(x)}$.
For the initial condition, expanding the generating function at large $x$ gives
\be
G_0(x) \simeq 1- \langle Z_0\rangle e^{-x}\quad \text{for} \quad x\to \infty. 
\ee
Thus, the initial condition effectively corresponds to a value of $\lambda = 1$ [see Eq.~\ref{eq:lambda-def}]. 
The selected value of $\lambda$ at late times is  the
\textit{minimum} between $\lambda^{\ast}$ and $1$. Borrowing terminology from a closely related problem of directed polymers on disordered trees \cite{Derrida_Polymers_1988},
we say that the system is in the glass class in cases where $\lambda$ converges to ${\lambda^\ast < 1}$, while cases where $\lambda^* > 1$ correspond to the paramagnetic class.\footnote{ The notion of a ``glass'' and a ``paramagnetic'' class comes from a mapping between linear recursion relations and the problem of a directed polymer on a disordered Cayley tree. The latter has a glass transition as a function of the ratio of temperature to disorder strength \cite{Derrida_Polymers_1988}. In the paramagnetic class the polymer has an entropy that is extensive in the depth of the tree, while in the glass class the entropy per unit length vanishes.
(We use the word ``class'' to avoid confusion with the entanglement ``phases''.
See  Ref.~\cite{Nahum_Measurement_2021}, Sec.~IV.G. for further discussion in the present context.)} 
In the present case --- the near-critical FMPT --- we have $\lambda^\ast<1$, as discussed above, so the solution with $\lambda^\ast$ is the relevant one (see Ref.~\cite{Nahum_Measurement_2021} for further discussion).

\subsection{Scaling behavior near the critical point}
\label{sec:force_scale}

We now consider scaling behavior of $Z_k$ near $\theta_c$. 
We focus on two specific quantities: the size of the order parameter
${\ztyp_\infty\equiv \lim_{k\rightarrow\infty} \ztyp_{k}}$ in the entangling phase as one closely approaches the transition  ($\theta < \theta_c$); 
and the dependence of $\ztyp_k$ on $k$ exactly at the critical point, ${\theta = \theta_c}$. 

Once we leave the disentangled phase, the linear recursion relation discussed above is no longer sufficient: the nonlinear terms in Eq.~\ref{eq:Zlinear} are essential to prevent $Z$ from diverging. 
However, previous work \cite{Nahum_Measurement_2021} showed that ultimately the critical scaling properties can be expressed in terms of the velocity function $v_\theta(\lambda)$ (Eq.~\ref{eq:force_velocity}). Here we briefly list the main results. (We will review the argument for these scaling forms in Sec.~\ref{sec:real} and App.~\ref{appendix:infinite_z}.)

Just on the entangling side of the critical point, 
\be
\ztyp_{\infty} \sim \exp{\left(-\frac{\pi}{\left|\text{Im}\, \lambda_{\theta}\right|}\right)}. \label{eq:infinit_z}
\ee
Here $\lambda_{\theta}$ is the solution of the equation $v^{(\lambda)}_{\theta} = 0$. Expanding Eq.~\ref{eq:force_velocity} around the critical point, we find
\be
\ztyp_{\infty}\sim \exp{\left(-\frac{C}{\sqrt{\theta_c-\theta}}\right)} \hspace{4mm} \text{for $\theta_c - \theta \ll 1$},
\label{eq:Ztyptheta_FMPT}
\ee
where $C$ is a constant coefficient. Using Eq.~\ref{eq:force_velocity} and Eq.~\ref{eq:infinit_z}, we can calculate the coefficient $C$,
\be
 C = \frac{\pi}{\sqrt{2}}\frac{\sqrt{\sum_i \langle A_i^{1/2}(\theta_c)(\ln A_i(\theta_c))^2\rangle}}{\sqrt{\left|\sum_i\langle \partial_{\theta}A^{1/2}_i(\theta)|_{\theta_c}\rangle\right|}}.
 \label{eq:CFMPT}
\ee
Numerically evaluating these averages gives $C = 2.903\pm 0.001$.
Taking appropriate derivatives  of Eq.~\ref{eqn:analytical_a}
(with respect to $\lambda$ to obtain the numerator of Eq.~\ref{eq:CFMPT} and with respect to $\theta$ to obtain the denominator) gives:
\be
C \simeq 2.90330810201297.
\ee
Exactly at the critical point, $\theta = \theta_c$, the velocity $v_\theta$ vanishes and the value of $\ln \ztyp_k$ evolves sub-ballistically with $k$.  Reference \cite{Nahum_Measurement_2021} argued that
\be
\ln \ztyp_k \sim -k^{1/3}. 
\label{eq:Ztypk}
\ee
Finally, we note that ${\lambda^\ast=1/2}$ can be viewed as a critical exponent determining the breadth of the probability distribution of $Z_k$ near criticality \cite{Nahum_Measurement_2021}. We discuss this further in Sec.~\ref{sec:compare}. 

Equations \ref{eq:thetacFMPT},~\ref{eq:Ztyptheta_FMPT} and \ref{eq:Ztypk} are our main predictions for the critical properties of the FMPT.
 We note that analogs of Eqs.~\ref{eq:Ztyptheta_FMPT} and \ref{eq:Ztypk} appear for the survival probability of a branching random walk with a moving absorbing wall \cite{derrida_survival_2007}, which can be understood in terms of an FKPP equation with a boundary condition that breaks translation symmetry.

\subsection{Numerical results}
\label{sec:FMPT_numerics}

In order to check the above predictions for $\theta_c$ and for the critical scaling, we perform numerical simulations of the collapse process.

A direct approach,  constructing many realizations of large trees, would be prohibitively expensive for large $k$ because of the exponential growth of the number of nodes with $k$. 
We overcome this limitation by using the pool method described in Refs.~\cite{Miller_Weak-disorder_1994,Monthus_Anderson_2009,Garcia-Mata_Scaling_2017}. 
Briefly, this method involves storing a large set (``pool'') $\{ Z_k^a \}_{a=1}^{N_\text{pool}}$ of values of $Z_k$ corresponding to trees of finite size $k$.
This pool is a way of approximately capturing the probability distribution $\mu_k$ of $Z_k$, via a large set of values drawn from this distribution, with
larger values of the pool size $N_\text{pool}$ giving a better approximation.
 A pool $\{ Z_{k+1} \}$ corresponding to trees of size $k+1$ can then be produced by randomly combining the values from the pool $\{ Z_k \}$ using  the process defined by Eq.~\ref{eq:final_density_matrix}.  We note that, while in the limit of $Z_k \ll 1$ one can use the linearized recursion relation of Eq.~\ref{eq:Zlinear}, throughout this paper our numerical data is obtained using the full recursion defined by Eq.~\ref{eq:final_density_matrix}, and does not assume small $Z_k$. Our numerical averages for $\ln \ztyp_k$ are taken over all elements of the pool. All our results in the main text are obtained with a pool size $10^6$, and we show in Appendix~\ref{appendix:poolsize} that the results we present are well-converged as a function of the pool size.

\begin{figure}[t]
    \centering
    \includegraphics[width = 1.0\columnwidth]{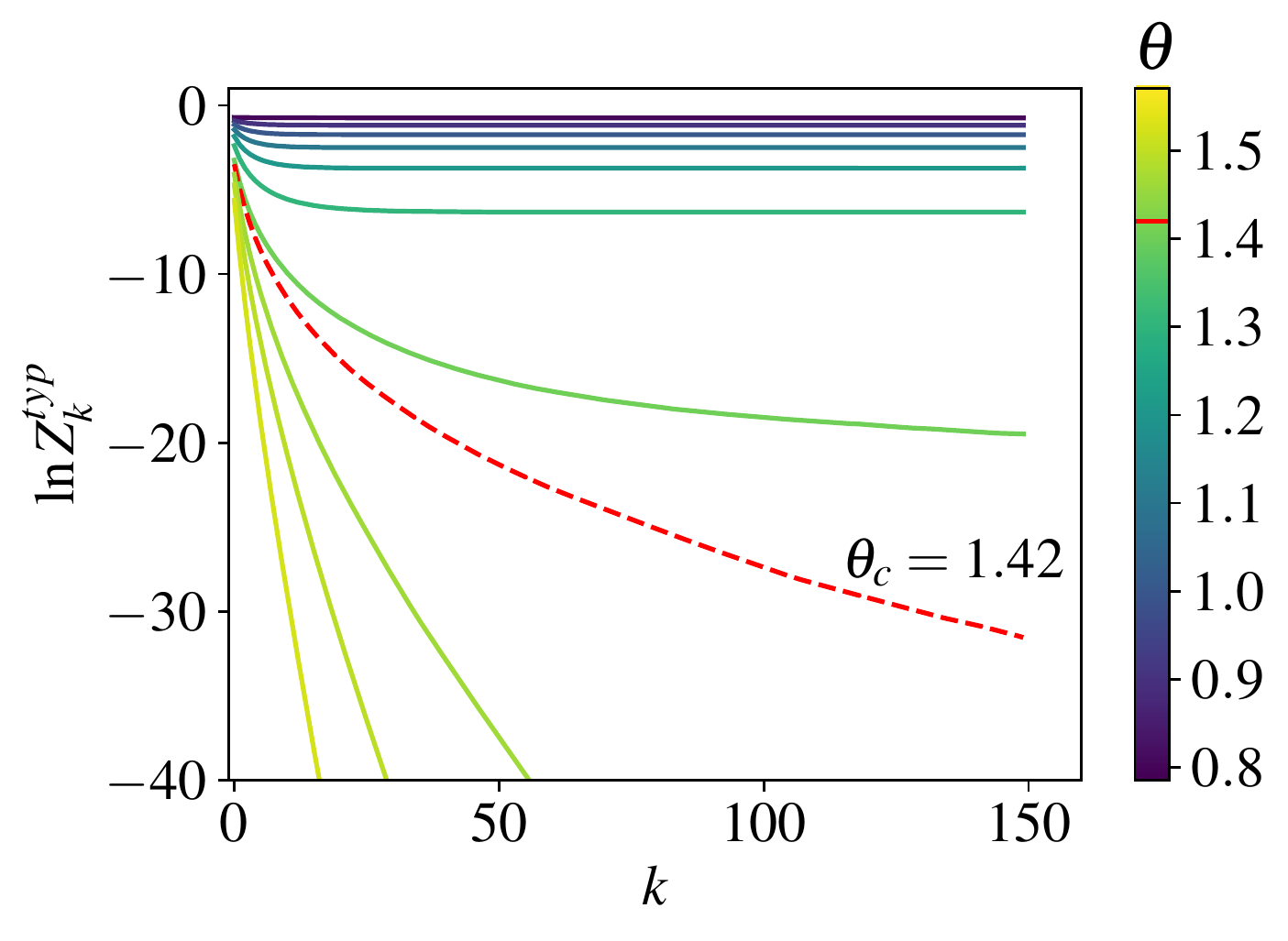}
    \caption{The evolution of $\ztyp$ with the tree depth $k$  for various $\theta$ in the forced measurement case. 
    The theoretical prediction for the critical measurement rate, $\theta_c = 1.42$, is indicated by the dashed line. 
    When the measurements are weak enough that ${\theta < \theta_c}$,  $\ztyp_k$ converges to a nonzero value in the limit ${k \rightarrow \infty}$. 
    When the measurements are stronger, ${\theta \geq \theta_c}$, $\ztyp_k$ tends to zero as $k \rightarrow \infty$, so that $\ln \ztyp$ tends to $-\infty$.}
    \label{fig:force_transition}
\end{figure}

Figure \ref{fig:force_transition} shows numerical results for $\ln \ztyp_k$ as a function of $k$ for different values of the measurement strength $\theta$. The existence of a critical measurement strength $\theta_c \approx 1.42$ can be clearly seen as a demarcation between two different classes of curves $\ln \ztyp_k$: those that saturate to a finite value at $k \rightarrow \infty$ and those that continue toward $-\infty$ (corresponding to $Z_k \rightarrow 0$).  
 
\begin{figure}[htb]
    \centering
    \includegraphics[width = 1.0 \columnwidth]{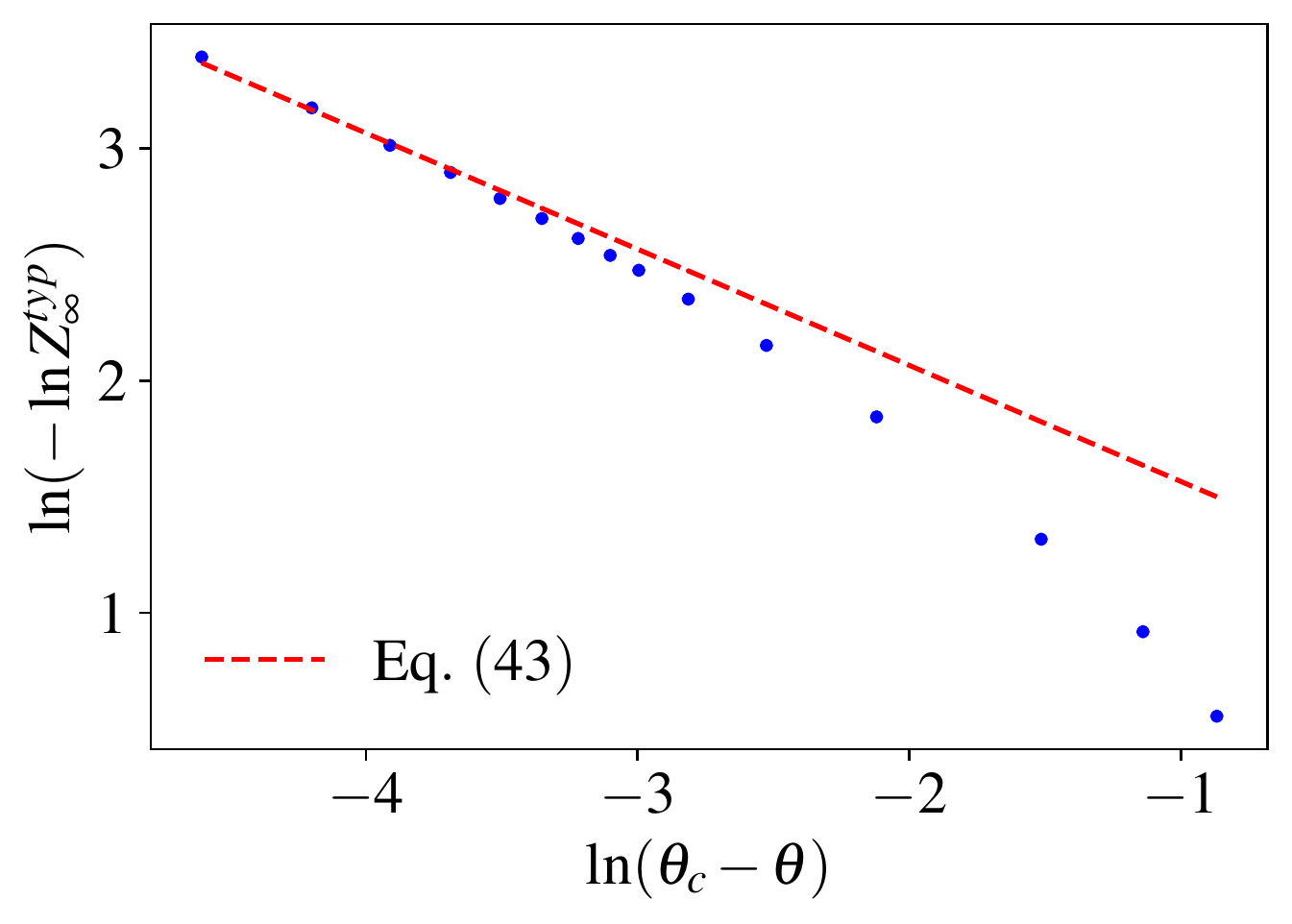}
    \caption{The scaling behavior of $\ztyp_{k \rightarrow \infty}$, as $\theta$ approaches the critical point from the entangling phase, is plotted for the case of forced measurements. The blue points show results from our numeric simulation, and the red dashed line shows the dependence predicted by Eq.~\ref{eq:Ztyptheta_FMPT} for the limit of asymptotically small $\theta_c - \theta$.}
    \label{fig:force_scale_zin}
\end{figure}

Figure~\ref{fig:force_scale_zin} shows numerical results for the scaling of the saturation value $\ln \ztyp_{k \rightarrow \infty}$ on the entangling side of the transition. The red dashed line indicates the theoretical prediction of Eq.~\ref{eq:Ztyptheta_FMPT}. 

\begin{figure}[t]
    \centering
    \includegraphics[width = 1.0\columnwidth]{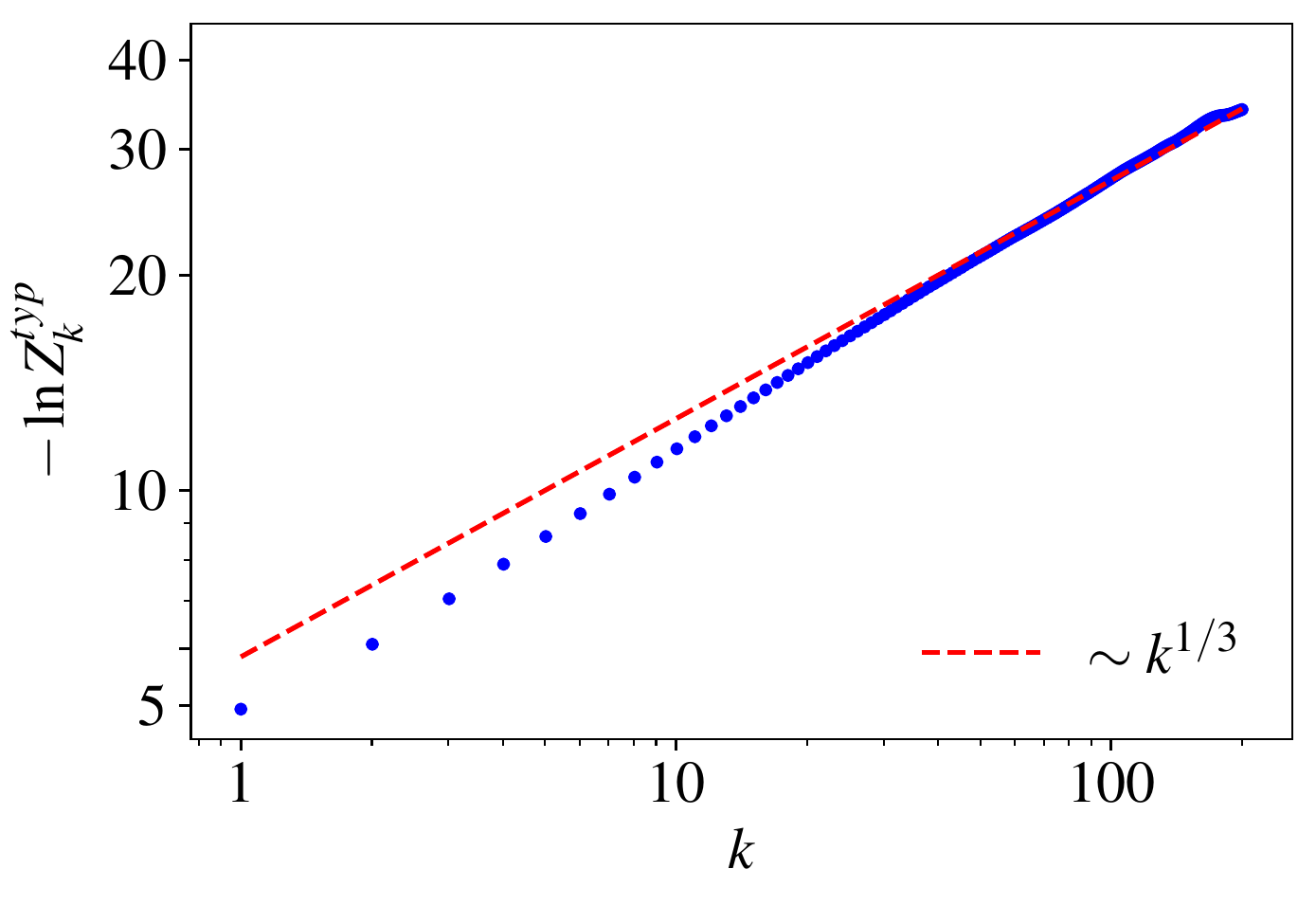}
    \caption{The scaling behavior of $\ln \ztyp_k$ exactly at the critical point is plotted for the case of forced measurements. Numerical results are shown as blue dots. 
   For comparison, the red dashed line shows the theoretical prediction ${-\ln \ztyp_k\sim k^{1/3}}$, which is a straight line in the log-log plot with an undetermined vertical offset.}
    \label{fig:force_scale_thetac}
\end{figure}

Figure~\ref{fig:force_scale_thetac} shows  the behavior of $\ztyp_k$ at the theoretically-predicted critical point, $\theta_c = 1.42$. The red dashed line shows the theoretical prediction of Eq.~\ref{eq:Ztypk}, $\ln \ztyp_k(\theta = \theta_c) \sim -k^{1/3}$.

\section{Real measurements}
\label{sec:real}

The preceding section considered the quantum tree with forced measurements. We found a FMPT with properties that are consistent with those found in  Ref.~\cite{Nahum_Measurement_2021}, although the details of the tree are slightly different. An interesting question, which is not directly considered in Ref.~\cite{Nahum_Measurement_2021}, is whether we can use a similar approach to also reveal the properties of the transition induced by real measurements. 

The key difference with the FMPT is that for real measurements the measurement outcomes $\sigma_1$, $\sigma_2$ and $\sigma'_2$ associated with a given node of the tree are no longer fixed, but are instead randomly chosen with probabilities determined by the Born rule
(leading to additional quantities to average over in each node).  
Since the outcomes of different measurements can be correlated,
the local three-index tensors $t=t(\sigma_1,\sigma_2,\sigma_2')$ that make up the network are no longer independently random objects: they now have nontrivial correlations not only within a node but also between different nodes.

We show in this section that averaging over the Born rule outcomes leads to different theoretical results for the MPT as compared to the FMPT.
Again we find that a travelling wave picture can be developed.
Notably, however, while the critical FMPT mapped to a travelling wave problem in the  ``glass'' class (with ${\lambda^\ast=1/2}$),
the critical MPT maps to the case 
${\lambda^\ast=1}$ which is at the boundary of the glass class with the paramagnetic class. 
The MPT still allows for two distinct dynamical phases, entangling and disentangling, separated by a critical measurement strength ${\theta = \theta_c}$ that is distinct from that of the FMPT case. 
The scaling  of $\ztyp_k$ as a function of $k$ and $\theta$ is qualitatively similar to that in the MPT, but there are universal differences between the problems that are reflected in the probability distribution of $Z_k$.

\subsection{Linearized recursion relation}
\label{sec:real_linear}

We wish to consider how the stochastic process defined by the node $t$ transforms the probability distribution $\mu_k(Z)$ into the probability distribution $\mu_{k+1}(Z)$.

For the recursive step, we consider (as in the previous section) a node ``event''
in which we start with known ``input'' density matrices $\rho_{k,1}$ and $\rho_{k,2}$ for the two spins.
By randomly choosing the necessary  Haar-random unitaries, 
and choosing the measurement outcomes with the Born-rule probabilities,
we obtain a random ``output'' density matrix $\rho_{k+1}$.

As in the previous section, it is sufficient to consider the case where the input  density matrices $\rho_{k,1}$ and $\rho_{k,2}$  have the diagonal form in Eq.~\ref{eq:rho}.
This simplification is possible because the probability distribution of the output density matrix, given the inputs, depends only on the eigenvalues of the inputs, and not on their eigenvectors.
This irrelevance of the eigenvectors follows from the Haar-randomness of the unitaries in the node, which means that there is no ``preferred basis'' for the spins.
(For a formal argument, see App.~\ref{app:gaugetransf}.)

Adopting this diagonal form, the input density matrices are evolved using the $t$-tensor  that defines the operation of a single node, 
which was given in Sec.~\ref{sec:collapseTN}.
Recalling these definitions,  
\be\label{eq:t3indexRepeated}
[t(\sigma_1, \sigma_2, \sigma_2')]^a_{cd}
=
\left[U \, (K^{(1)}_{\sigma_1}\otimes K^{(2)}_{\sigma_2})\right]^{a \sigma_2'}_{cd},
\ee
This tensor $t=t(\sigma_1,\sigma_2,\sigma_2')$ depends on the 2-site unitary and on the random bases for the measurements (this dependence will be left implicit), and also on the measurement outcomes (these will sometimes be shown as explicit arguments).
As before, the output density matrix is given by 
\be
\rho_f = \f{t (\rho_1\otimes \rho_2) t^\dag}{\tr \, t (\rho_1\otimes \rho_2) t^\dag }. 
\ee
The measurement outcomes are chosen probabilistically with the Born rule:
for given initial states and unitaries, 
the probability of a set of outcomes $(\sigma_1, \sigma_2, \sigma_2')$ may be written 
\be\label{eq:p_real_lin_sec}
p(\sigma_1, \sigma_2, \sigma_2') = \tr \, t(\sigma_1,\sigma_2,\sigma_2') (\rho_1\otimes \rho_2) t(\sigma_1,\sigma_2,\sigma_2')^\dag,
\ee
as discussed in Sec.~\ref{sec:collapse}.
Below we also write this probability as $p(s)$, using ${s=(\sigma_1,\sigma_2,\sigma'_2)}$ to denote the collection of measurement outcomes for the node. For notational simplicity we have suppressed the dependence of $p(s)$ on the initial states and on the random unitaries.

Formally, the value of $Z_{k+1}$ is a function of
the initial states $Z_{k,1}$, $Z_{k,2}$
 and of the $t$-tensor (and through it the measurement outcomes):
\be
Z_{k+1} = f\lf t( \sigma_1,\sigma_2,\sigma'_2), Z_k^1,Z_k^2 \ri. 
\ee 
For a fixed set of measurement outcomes, the expansion of the function $f$ in terms of the arguments $Z_{k,1}, Z_{k,2}$ proceeds identically to the forced measurement case (Sec.~\ref{sec:force_linear}), and thus Eqs.~\ref{linear_coeff1} and~\ref{linear_coeff2} for the coefficients $A_1$ and $A_2$ are still valid for small $Z_{k,1}$, $Z_{k,2}$:
\begin{align}
    Z_{k+1} &\simeq  A_1(\sigma_1,\sigma_2,\sigma'_2) Z_{k,1} +A_2(\sigma_1,\sigma_2,\sigma'_2)Z_{k,2}, 
\label{real_lin_recursion}
\end{align}
with
\begin{align}
    A_1 &= \frac{\left|t_{11}^{1}t_{21}^{2}-t_{21}^{1}t_{11}^{2}\right|^2}{\left(|t_{11}^{1}|^2+|t_{11}^{2}|^2\right)^2}, &
    A_2 &=\frac{\left|t_{11}^{1}t_{12}^{2}-t_{12}^{1}t_{11}^{2}\right|^2}{\left(|t_{11}^{1}|^2+|t_{11}^{2}|^2\right)^2},
    \label{real_lin_coeff2}
\end{align}
where the coefficients ${A_1=A_1(\sigma_1, \sigma_2, \sigma_2')}$ and ${A_2=A_2(\sigma_1, \sigma_2, \sigma_2')}$ are now dependent on the measurement outcomes.

This dependence on the measurement outcomes means that, unlike the forced measurement case, 
the coefficients $A_i$ in Eq.~\ref{real_lin_recursion} are not chosen independently of the $Z_{k,i}$, since the 
probability of given measurement outcomes is itself dependent on  $Z_{k,1}$ and $Z_{k,2}$.
Fortunately, however, do not need to consider the full dependence.
In particular, in the disentangling phase, where $Z_k$ becomes arbitrarily small at large $k$, we can approximate the measurement probabilities by their $Z\rightarrow 0$ limits. 
This approximation is sufficient to locate the transition.

Expanding Eq.~\ref{eq:p_real_lin_sec} in $Z_{k,1}$, $Z_{k,2}$ and retaining only the zeroth-order term, and using the fact that $\rho_{1,2}$ are diagonal  in our chosen basis and have the form in Eq.~\ref{eq:rho}, we find 
\be
p(\sigma_1, \sigma_2, \sigma_2') \simeq
|t_{11}^1|^2 + |t_{11}^2|^2.
\ee
Using the expressions for $t$ in Eqs.~\ref{tilde-t-tensor},~\ref{eq:t3index},
 and the property of the Kraus operators
\be
\sum_\sigma K_\sigma^\dag K_\sigma = \mathbf{1},
\ee
one can check that the above probability ${p}$ is normalized, such that $\sum_{\sigma_1,\sigma_2,\sigma'_2} {p}(\sigma_1,\sigma_2,\sigma'_2) = 1$.

Using Eqs.~\ref{real_lin_recursion}-\ref{real_lin_coeff2}, the generating function $G_k(x)$ has the recursion relation
\be
G_{k+1}(x) =\sum_{s} \langle p(s)G_k( x-\ln A_1(s,\theta) )
G_k(x-\ln A_2(s,\theta)) \rangle. 
\label{eq:recursionGGG}
\ee
Here we use $s$ to represent the set of all three measurement outcomes. The angle brackets are the average over the Haar-random unitary operations: the two-site random unitary $U$, and the 1-site random unitaries in $K^{(i)}$ which set the bases for the weak measurements. The argument $\theta$ is the strength of weak measurements. 

As before, at late times $G_k(x)$ converges to a traveling wave solution, parameterized by a variable $\lambda$ which must be determined.
The solution for a given $\lambda$ has velocity
\be
v_{\theta}(\lambda) = \frac{1}{\lambda} \ln \sum_{s}\left\langle p(s)(A^{\lambda}_1(s,\theta)+A^{\lambda}_2(s,\theta)) \right\rangle. \label{real_velocity}
\ee
For a given $\theta$, there is a special value $\lambda=\lambda^*$ that gives the minimal velocity,
\be
\left.\frac{\partial v_\theta(\lambda)}{\partial \lambda}\right|_{\lambda =\lambda^{\ast}} = 0.\label{eqn:velocity_derivative_real}
\ee
The velocity selected is given by 
\be
v_\theta =\left\{
\begin{array}{ll}
     v_\theta(\lambda^\ast) & \text{if $\lambda^\ast\leq 1$}   \\
      v_\theta(1) & \text{if $\lambda^\ast\geq 1$}
\end{array}
\right. ,
\ee
and the critical point $\theta_c$ 
is defined by the vanishing of this velocity. 
Recall the physical meaning of this velocity: in the disentangling phase, $v_\theta$ is negative, and ${\ztyp_k\sim e^{-|v_\theta| k}}$.

If $\lambda^\ast\leq 1$, the critical point is given by solving the following equations for the unknowns $\lambda^\ast$ and $\theta_c$:
\begin{align}
    v_{\theta_c}(\lambda^{\ast}) & = 0,
   &
    \partial_{\lambda^\ast} v_{\theta_c} (\lambda^\ast) & = 0.
\end{align}
These equations may be rewritten as 
\begin{align}
\sum_{s}  \sum_{i=1,2} 
\left\langle p(s) A_i(s,\theta_c)^{\lambda^{\ast}}\right\rangle & =1,
\label{eq:vzeroMPT} \\
\sum_{s} \sum_{i=1,2}  \left\langle p(s)   A_i(s,\theta_c)^{\lambda^{\ast}}
\ln A_i(s,\theta_c) \right\rangle &  = 0. 
\label{eq:dvdlambdaMPT}
\end{align}

In Appendix~\ref{appendix:lambda} we show that the coefficients $A_1(s,\theta)$, $A_2(s, \theta)$ satisfy the striking identity
\be
\sum_{s} \left\langle p(s) A_i(s,\theta) \ln A_i(s,\theta)\right\rangle =0 
\label{real_identity}
\ee
for both $i = 1$ and $i=2$. 
As a result, Eq.~\ref{eq:dvdlambdaMPT} is satisfied by $\lambda^\ast=1$.
Equation \ref{eq:vzeroMPT} then gives the value of $\theta_c$:
\be\label{eq:eqnformptthetac}
\sum_{s}  \sum_{i=1,2} 
\left\langle p(s) A_i(s,\theta_c) \right\rangle  =1.
\ee
This equation may also be written
\be\label{eq:mptcriticalpoint2}
16 \left\langle 
p(\uparrow,\uparrow,\uparrow) 
A_1((\uparrow,\uparrow,\uparrow),\theta_c) \right \rangle = 1,
\ee
since the average in Eq.~\ref{eq:eqnformptthetac} is independent of $s$ (which runs over 8 possible measurement sequences) and of $i$.

Thus, at the critical point the wavefront solution  has $\lambda^{\ast} = 1$, which means that the linear recursion relation is poised exactly at the boundary point between the glass and paramagnetic classes (which yield different critical scaling once nonlinearities are included
\cite{Nahum_Measurement_2021}).  Note that $\lambda^\ast = 1$ only at the critical point, and in general the value of $\lambda^{\ast}$ varies with $\theta$ near the critical point as $\lambda^\ast - 1 \propto  \theta_c-\theta$.

Numerically integrating the left-hand side of Eq.~\ref{eq:eqnformptthetac}
allows us to estimate the critical measurement strength $\theta_c$. This process gives ${\theta_c = 1.100 \pm 0.001}$. 

The above procedure, and the critical point equation (\ref{eq:eqnformptthetac}), applies for a larger class of tree models with measurements,  in which the averages $\langle\ldots \rangle$ have the unitary invariance property discussed in Sec.~\ref{sec:force_linear}. 
Surprisingly, the simple structure of the node tensor $t$ in the particular model under study  allows the integral to be performed analytically.
Any average of the form ${\langle p^\kappa A_i^\lambda\rangle}$ can be evaluated explicitly as a function of $\theta$ (App.~\ref{app:averages}),
in particular
\begin{align} \notag
& \langle p(s) A_i^\lambda\rangle = \\
& \, \, \f{
3 \times 2^{1-2\lambda} (\sin 2\theta)^{2\lambda} (\cos 2\theta)^{-1} 
\lf 
(\sin \theta)^{4-4\lambda} - (\cos \theta)^{4-4\lambda} 
\ri
}{
(\lambda+1)(\lambda+2)(\lambda-3)(\lambda-4)(2\lambda-2)
} \notag
\end{align}
The critical point equation (\ref{eq:mptcriticalpoint2}) then becomes
\begin{align}
\f{\ln \gamma}{\gamma^2 - 1/\gamma^2 } & = \f{3}{16}, &
&\text{for $\gamma=\tan\theta_c$},
\end{align}
which has the solution
\be
\theta_c \simeq 1.10010302468401.
\ee

\subsection{Scaling behavior near the critical point}
\label{sec:real_scaling}

The primary qualitative difference between the real measurement and forced measurement critical points is that the real measurement case has $\lambda^* = 1$,
while the forced measurement case has ${\lambda^*=1/2}$.
Recursion relations with  ${\lambda^*<1}$ and ${\lambda^*>1}$ generally exhibit different order parameter scaling, so the scaling for the MPT in our model is a subtle question.\footnote{The fact that the distributions of the $A_i$ in Eq.~\ref{real_lin_recursion} are dependent on the $Z_k$ is also a new feature compared to the previously studied cases.} Below we give a heuristic argument fixing the leading scaling, leaving a more detailed and rigorous study of the recursion relations to the future. 

Slightly inside the weak-measurement phase, we find that
\begin{align}
 \ztyp_{\infty} & \sim \exp{\left( -\frac{C'}{\sqrt{\theta_c-\theta}} \right)} &   (\text{for } \theta_c - \theta \ll 1)
\label{eq:ztypthetaMPT}
\end{align}
where the constant $C'$ is discussed below. Exactly at the critical point, 
\begin{align}
 \ln \ztyp_k & \sim - k^{1/3} &  (\text{for } \theta & = \theta_c). 
\label{eq:ztypk_MPT}
\end{align}
These scaling forms are compared with simulation data in Sec.~\ref{sec:real_pt}.
They are similar to those for forced measurements (Sec.~\ref{sec:force}). 
On the other hand, $\lambda^*$  can be viewed as a critical exponent (governing the breadth of the critical probability distribution for $Z_k$) that distinguishes the real and forced measurement critical points. This difference is discussed in Sec.~\ref{sec:compare}.

Now we discuss the order parameter scaling in slightly more detail. (Some readers may prefer to skip the remainder of this subsection.)

It is convenient to define 
\be
H(x) =\lim_{k\rightarrow\infty} 1-G_k(x),
\ee
which is nontrivial in the entangling phase.
In App.~\ref{appendix:infinite_z} we give an exact linear equation for $H(x)$ that holds for ${x\gg \ln Z_\text{typ}}$, i.e. far to the right of the front. 
This equation simplifies further when  $x$ is also sufficiently large and negative. In this intermediate regime
(loosely speaking, negative $x$ with 
$1\ll |x| \ll |\ln \ztyp|$),
the analysis is similar to that in the previous section, where we may take  ${H(x)\sim e^{-\lambda x}}$. 
{For the critical scaling on the entangling side of the transition,} however, we are looking for a solution that is \textit{stationary} (independent of $k$), so the 
parameter $\lambda$ must be chosen  such that the traveling wave velocity $v_\theta(\lambda)$ vanishes. 
The key point is that satisfying this condition requires a complex $\lambda$ \cite{brunet1997shift}:
\be
\lambda = 1\pm i c \sqrt{\theta_c-\theta} + O(\theta_c-\theta).
\ee
(Here $c$ is a constant determined by  Eq.~\ref{eq:vzeroMPT}.) Taking a real combination of the two complex solutions then gives
\be\label{eq:Hfmla}
H(x) \propto e^{-x} \sin \lf \phi + c
\sqrt{\theta_c-\theta} \, x \ri, 
\ee
where $\phi$ is an undetermined constant.

By considering how this solution in the intermediate regime ${1\ll |x| \ll |\ln \ztyp|}$ matches onto solutions to the left and right of this interval we may fix the scaling of  $\ztyp$.
Note that Eq.~\ref{eq:Hfmla} gives a logarithmic ``slope'' that is close to $-1$ for most values of $x$:
\be
\partial_x \ln H = - 1 + \f{c \sqrt{\theta_c-\theta}}{\tan (\phi + c
\sqrt{\theta_c-\theta} \, x) }
\label{eq:slopeeqn}
\ee
The exceptions are close to the zeroes  of the tangent, for which the second term is no longer small.

As $x$ approaches the left hand side of the regime where (\ref{eq:slopeeqn}) is valid, i.e. as ${x}$ approaches ${\ln \ztyp}$, we expect that the magnitude ${| \partial_x \ln H |}$ of the slope decreases  towards zero, in order to match the plateau in $H$ for ${x\ll \ln \ztyp}$.
This condition suggests that the argument of the $\tan$ function should vanish at a value of $x$ close to $\ln \ztyp$ \cite{Nahum_Measurement_2021}.
Consequently, 
\be\label{eq:phisqrt}
\ln \ztyp \sim  - \f{C'}{\sqrt{\theta_c-\theta}},
\ee
with $C'=  \phi /c$.
In principle, we should now use the matching condition on the other side of the intermediate regime (where $x$ approaches $-1$) to fix the numerical value of $\phi$. 
We suspect that the value ${\phi=\pi/2}$ is required, in order for Eq.~\ref{eq:slopeeqn} to be consistent with an expansion of the generating function in terms of moments of $Z_k$  (see App.~\ref{appendix:infinite_z}). However, confirming this would require a more careful analysis of corrections to Eq.~\ref{eq:slopeeqn}.

Finally, given Eq.~\ref{eq:Hfmla}, the form (\ref{eq:ztypk_MPT}) for the $k$-dependence of $\ztyp_k$ exactly at the critical point follows from the same heuristic argument as in the forced measurement case.\footnote{See Sec.~IV.H.4 of Ref.~\cite{Nahum_Measurement_2021}.}

\subsection{Numerical results}
\label{sec:real_pt}

We can  confirm our analytical results for the MPT numerically, using the same pool  algorithm described in Sec.~\ref{sec:FMPT_numerics} for the FMPT. The only difference for real measurements is that, in applying the recursion relation defined by Eq.~\ref{eq:final_density_matrix}, one must select the measurement outcomes with the Born-rule probabilities in~Eq.~\ref{eq:real_outcomep}.

\begin{figure}[t]
    \centering
    \includegraphics[width = 1.0\columnwidth]{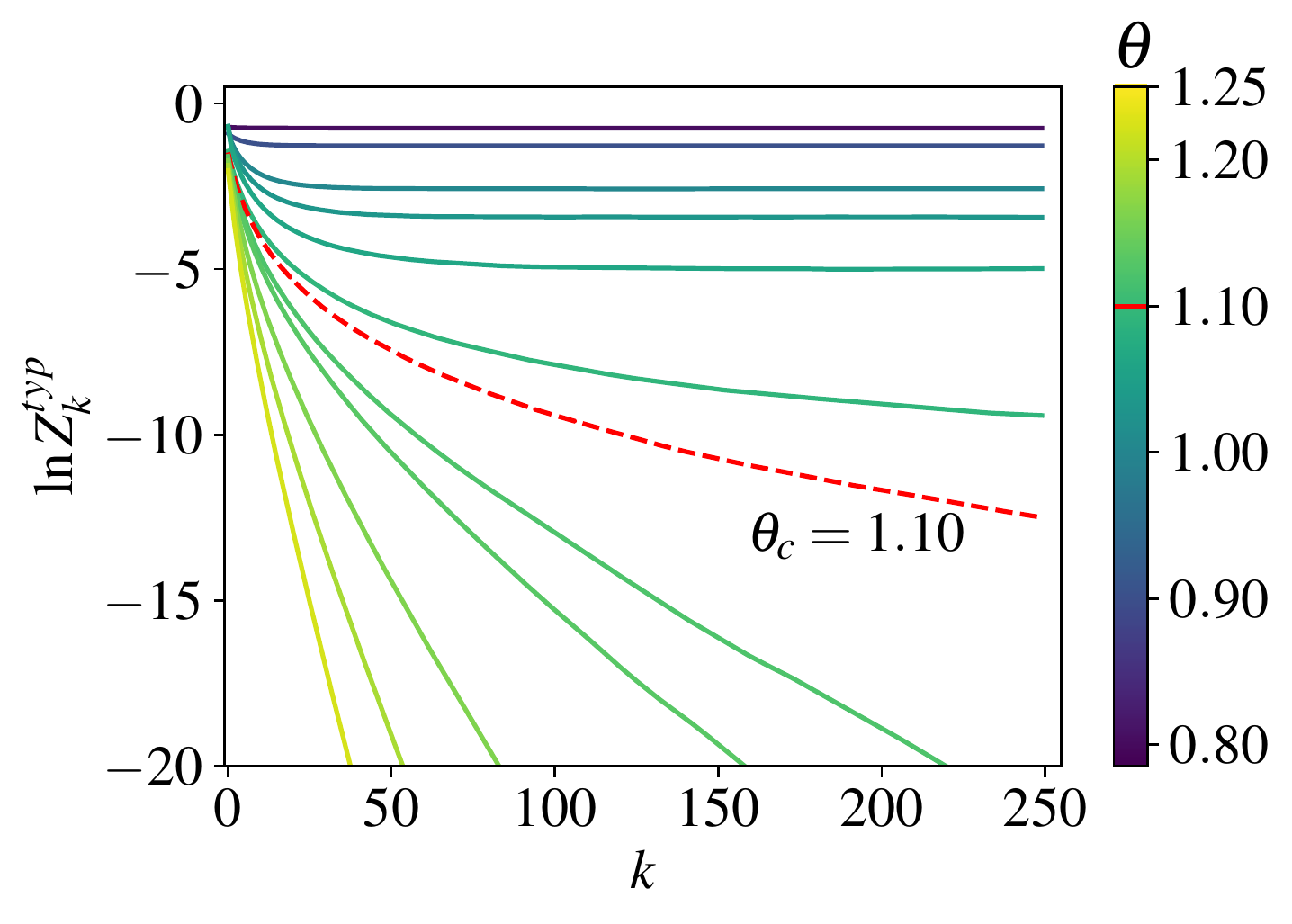}
    \caption{The evolution of $\ztyp$ with the depth $k$ of the tree for the case of real measurements, plotted for different values of the measurement strength $\theta$. The red dashed line represents the value of $\theta$ corresponding to the theoretically predicted critical point ($\theta_c \approx 1.10$).  The entangling phase ($\theta < \theta_c$) is characterized by a nonzero value of $\ztyp_{k \rightarrow \infty}$, and the disentangling phase ($\theta > \theta_c$) has ${\ztyp_{k\rightarrow \infty}= 0}$, i.e. ${\ln \ztyp_{k\rightarrow \infty}= -\infty}$.}
    \label{fig:real_transition}
\end{figure}

Figure~\ref{fig:real_transition} shows the evolution of $\ln \ztyp_k$ with the tree depth $k$ for different values of the measurement strength $\theta$. As in the forced measurement case, one can see two phases separated by the critical measurement strength $\theta_c$. In the entangling phase, ${\theta < \theta_c}$, the value of $\ln \ztyp_k$ is finite in the limit $k \rightarrow \infty$, so that an initially mixed state remains mixed regardless of the size of the tree. 
On the other hand, in the disentangling phase, $\theta > \theta_c$, the value of $\ln \ztyp_k$ is $-\infty$ in the limit $k \rightarrow \infty$, which means that an initially mixed state is completely purified in the limit of a large tree. The red curve in the plot represents the critical value $\theta_c$ obtained by the linear recursion relation; one can see that this value is consistent with the numerical simulation.

\begin{figure}[t]
    \centering
    \includegraphics[width = 1.0\columnwidth]{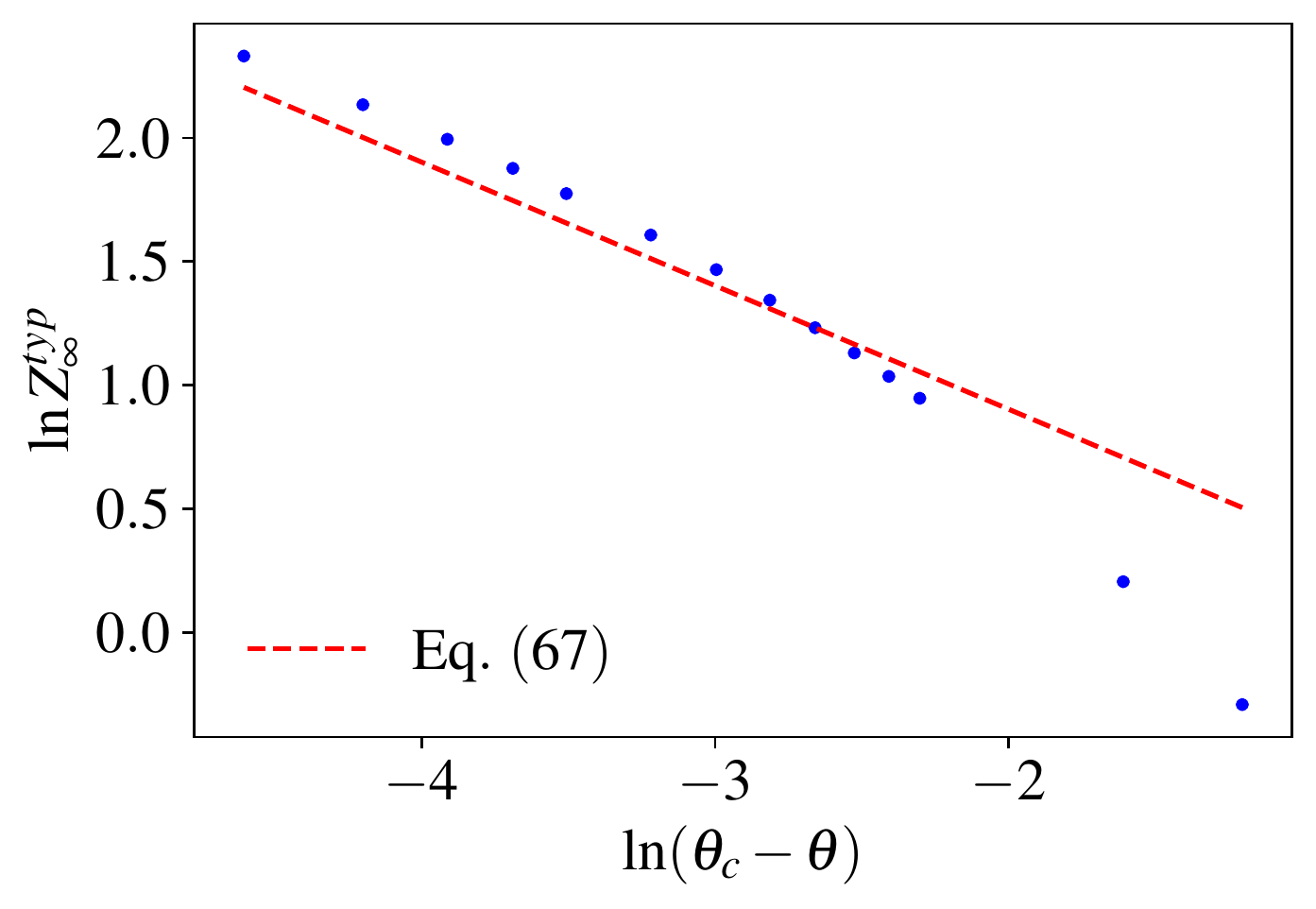}
    \caption{The scaling behavior of $\ztyp_{k \rightarrow \infty}$ as $\theta$ approaches the critical value from the entangling side of the transition is plotted for the case of real measurements. The blue points show results from our numeric simulation and the red dashed line shows the functional dependence predicted by Eq.~\ref{eq:ztypthetaMPT} for the limit of asymptotically small $\theta_c - \theta$. The value of the parameter $C'$ is calculated with $\phi = \pi/2$.}
    \label{fig:real_scale_zin}
\end{figure}

\begin{figure}[t]
    \centering
    \includegraphics[width = 1.0\columnwidth]{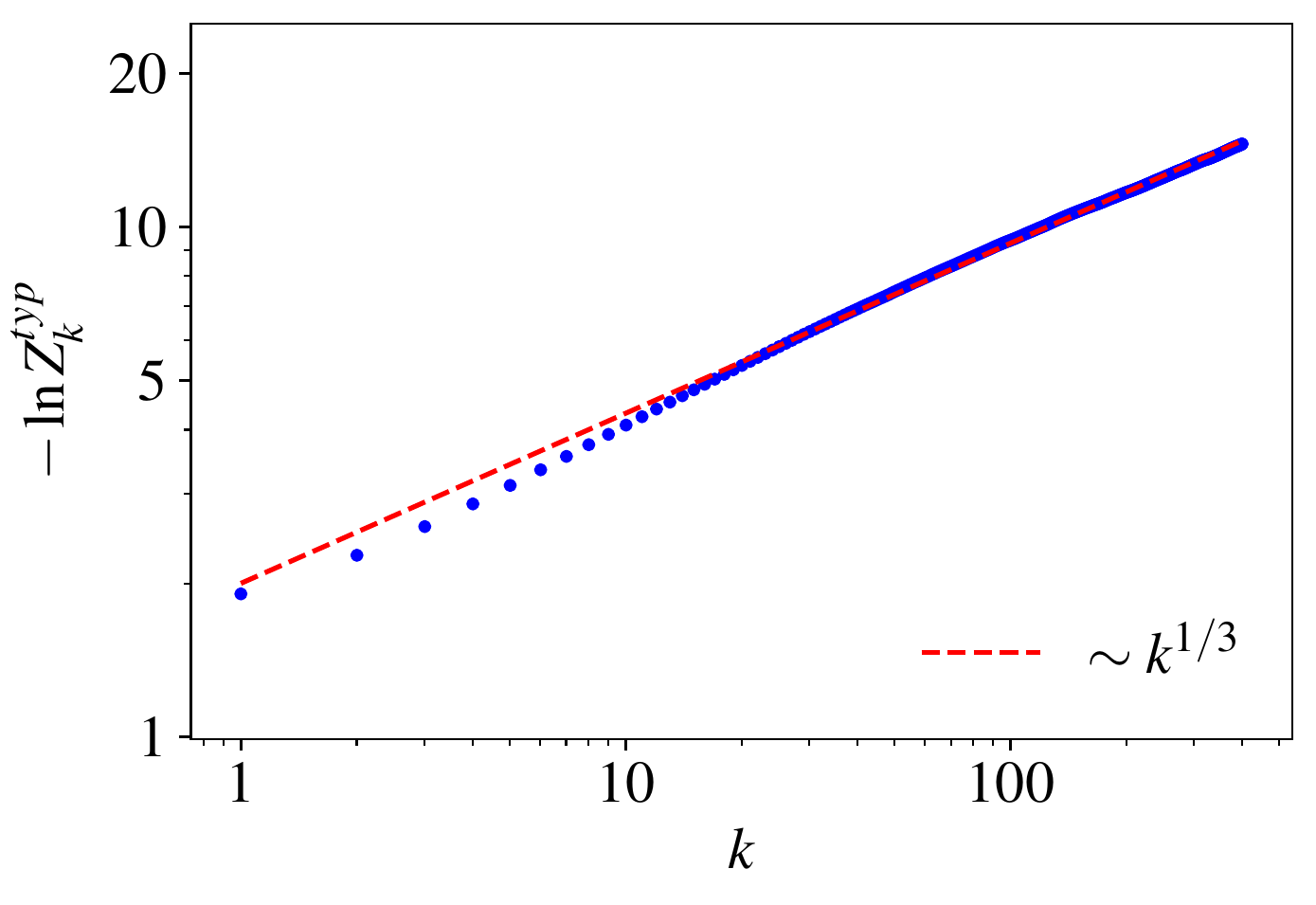}
    \caption{The value of $\ln \ztyp_k$ at the critical point ${\theta = \theta_c \approx 1.10}$ is plotted for the case of real measurements. The red dashed line shows the dependence $\ln \ztyp_k \sim -k^{1/3}$ predicted by Eq.~\ref{eq:ztypk_MPT}. }
    \label{fig:real_scale_thetac}
\end{figure}

Figure~\ref{fig:real_scale_zin} shows the scaling behavior of $\ln \ztyp_{k \rightarrow \infty}$ as one approaches the transition from the entangling side. The red dashed line shows the predicted dependence given by Eq.~\ref{eq:ztypthetaMPT}.  The predicted scaling of $\ln \ztyp_k$ with $k$ at $\theta = \theta_c$ is also confirmed in Fig.~\ref{fig:real_scale_thetac}, which shows $\ln \ztyp_k \sim - k^{1/3}$.


\subsection{Comparison: real vs. forced measurements}
\label{sec:compare}

In closing this section we briefly comment on the differences between the forced measurement and real measurement cases. In both cases the system exhibits a phase transition between entangling and disentangling dynamical phases as a function of increasing measurement strength. However, the real measurement case has a smaller critical value of $\theta$, which means that the entangling phase is smaller for real measurements than for forced measurements.  

This difference between the two cases implies that real measurements are more strongly purifying than forced measurements for the present dynamics. 
To get some intuition for the difference between forced and real measurements in terms of purification, one can consider the simple example of a single weak measurement on a single spin. 
Specifically, we compare the effect of a true measurement to the effect of a forced measurement in which the outcome is chosen independently of the state and  with 50\% probability for $\uparrow$ or $\downarrow$.

Let us visualize the spin's initial density matrix $\rho_\text{init}$ as a point in the interior of the Bloch sphere (recall that pure states lie on the surface of the Bloch sphere, while mixed states lie in the interior).
Without loss of generality,  assume that this initial density matrix is closer to the north pole ($\ket{\uparrow}$) than to the south pole ($\ket{\downarrow}$).

Now make a very weak \cite{tilloy2015spikes} measurement (${\theta= \pi/4+\epsilon}$) in the $S_z$ basis.
After the measurement, the density matrix is altered so that the corresponding point in the Bloch sphere is displaced very slightly. 
If the measurement outcome is $\uparrow$, the displacement takes the point closer to  
the north pole.
Since the state is initially  closer to the north pole, this displacement also brings it closer to the surface of the Bloch sphere, i.e.\ it is purifying.
By contrast, if the measurement outcome is $\downarrow$, the density matrix gets closer to the south pole and becomes less pure. 

Real measurements make the first outcome (the purifying one) more likely, while for forced measurements the two outcomes are equally likely. The result is that, after averaging over outcomes, the real measurement gives a mean increase in purity of order $\epsilon$, while the forced measurement gives a mean change that  is higher-order in $\epsilon$ 
(we omit the detailed calculation).

\begin{figure}[t]
    \centering
    \includegraphics[width = 0.9\columnwidth]{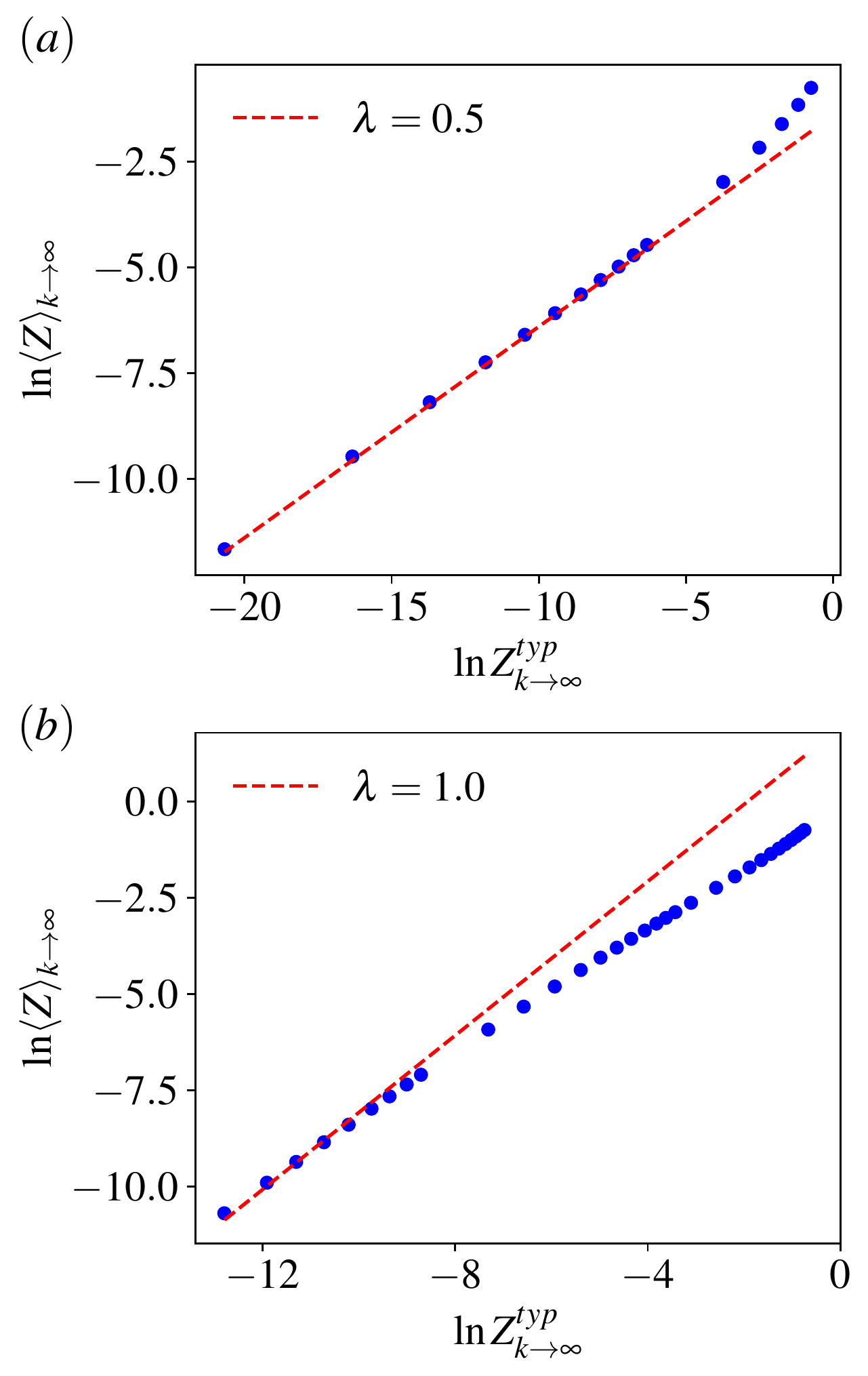}
    \caption{The relationship between the typical value of $Z$ (defined by Eq.~\ref{eq:ztyp}) and the average value of $Z$ depends on the parameter $\lambda$.  The expected relationship is  $\ln \langle Z \rangle\sim \lambda \ln\ztyp$,
    with $\lambda=1/2$ for forced measurements and 
    $\lambda=1$ for true measurements. 
    Plot (a) compares this prediction (red dashed line) with data (blue dots) for forced measurements. Plot (b) shows the case of true measurements. The somewhat worse agreement for case (b) is likely due to a logarithmic prefactor that we have neglected in the relation between $\ztyp$ and $\langle Z\rangle$.
    }
 \label{fig:avg_typ_compare}
\end{figure}

The other prominent difference between the MPT and FMPT is the different value of the critical exponent $\lambda$ that characterizes the two transitions; $\lambda = 1/2$ for the FMPT, which is within the glass class, while $\lambda = 1$ for the MPT, which is at the boundary between the glass and paramagnetic classes. Ultimately this difference arises because the Born rule probability $p(s)$ depends on the set of measurement outcomes $s$.
This dependence implies that the nodes of the tensor network $T$ are no longer independent. 
The measurement outcomes at a node, which go into determining the node tensor
${t(\sigma_1,\sigma_2,\sigma'_2)}$,
have probabilities that depend on the density matrices $\rho_1$, $\rho_2$ coming from the earlier part of the tree.\footnote{In our treatment of the node
we considered (without loss of generality) the case where $\rho_1$ and $\rho_2$ are diagonal. The correlations between $\rho_i$ and $t$ then imply that  the resulting distribution for the node tensor $t$ is not invariant under the rotations
${t^{a}_{bc}\to t^{a}_{b'c}u_{b'b}}$ or ${t^{a}_{bc}\to t^{a}_{bc'}u_{c'c}}$.
That is, $\rho_{1,2}$ have picked out preferred states on the bonds, and the distribution of $t$ is sensitive to these preferred states.}

Nonetheless, the Haar-randomness of the 2-site unitaries means that the ensemble of density matrices  $\rho_k$ for a tree of $k$ generations can be characterized only by a distribution of singular values $Z_k$. This allows us to treat the problem  via a recurrence relation for $Z_k$.

While this difference in the value of $\lambda$ is not reflected in the critical scaling of $\ln \ztyp_k$ near the transition (compare Eqs.~\ref{eq:Ztyptheta_FMPT} and \ref{eq:Ztypk} with Eqs.~\ref{eq:ztypthetaMPT} and \ref{eq:ztypk_MPT}), it does have an observable consequence in terms of the distribution of values of $Z_k$ among different realizations of the tensor network. 
Specifically, the exponent $\lambda$ controls the probability density ${P(\ln Z)\,d\ln Z}$
for $\ln Z$ \cite{Derrida_Polymers_1988, Nahum_Measurement_2021}.
We consider the regime just inside the entangling phase, with finite asymptotic values ${\ztyp=\ztyp_{k\rightarrow\infty}}$ and ${\langle Z\rangle = \langle Z_{k}\rangle_{k\rightarrow\infty} }$
(similar considerations apply just on the other side of the transition, and at the critical point, for the probability distribution at large finite $k$).
When $\ztyp$ is very small, there is a wide range for which
${1 \ll |\ln Z| \ll |\ln \ztyp|}$, and in this range
\be
{P(\ln Z) \propto (\ztyp/Z)^\lambda}
\ee
when $\lambda\leq 1$
(neglecting a possible more slowly-varying prefactor).
Consequently,
when $\ztyp$ is small the relationship between 
$\langle Z \rangle$ and $\ztyp$ is 
{$\langle Z \rangle\sim \left(\ztyp \right)^{\lambda}$}
(again, neglecting a possible more slowly-varying prefactor).
This relationship is confirmed by our simulation data in Fig.~\ref{fig:avg_typ_compare} for both the  forced measurement and real measurement cases. Specifically, the two cases exhibit a different slope $\lambda$ when  $\ln\langle Z_{k\to\infty}\rangle$ is plotted against $\ln \
\ztyp_{k\to\infty}$. This plot provides an independent confirmation of the different value of $\lambda$ between the two cases.

\section{The expansion process and experimental protocols}
\label{sec:expansion}

Having discussed the collapse process in the previous sections, we now turn to the phase transition in the expansion process, and we suggest a schematic protocol for detecting it experimentally.

\begin{figure}[t]
    \centering
    \includegraphics[width = 0.8\columnwidth]{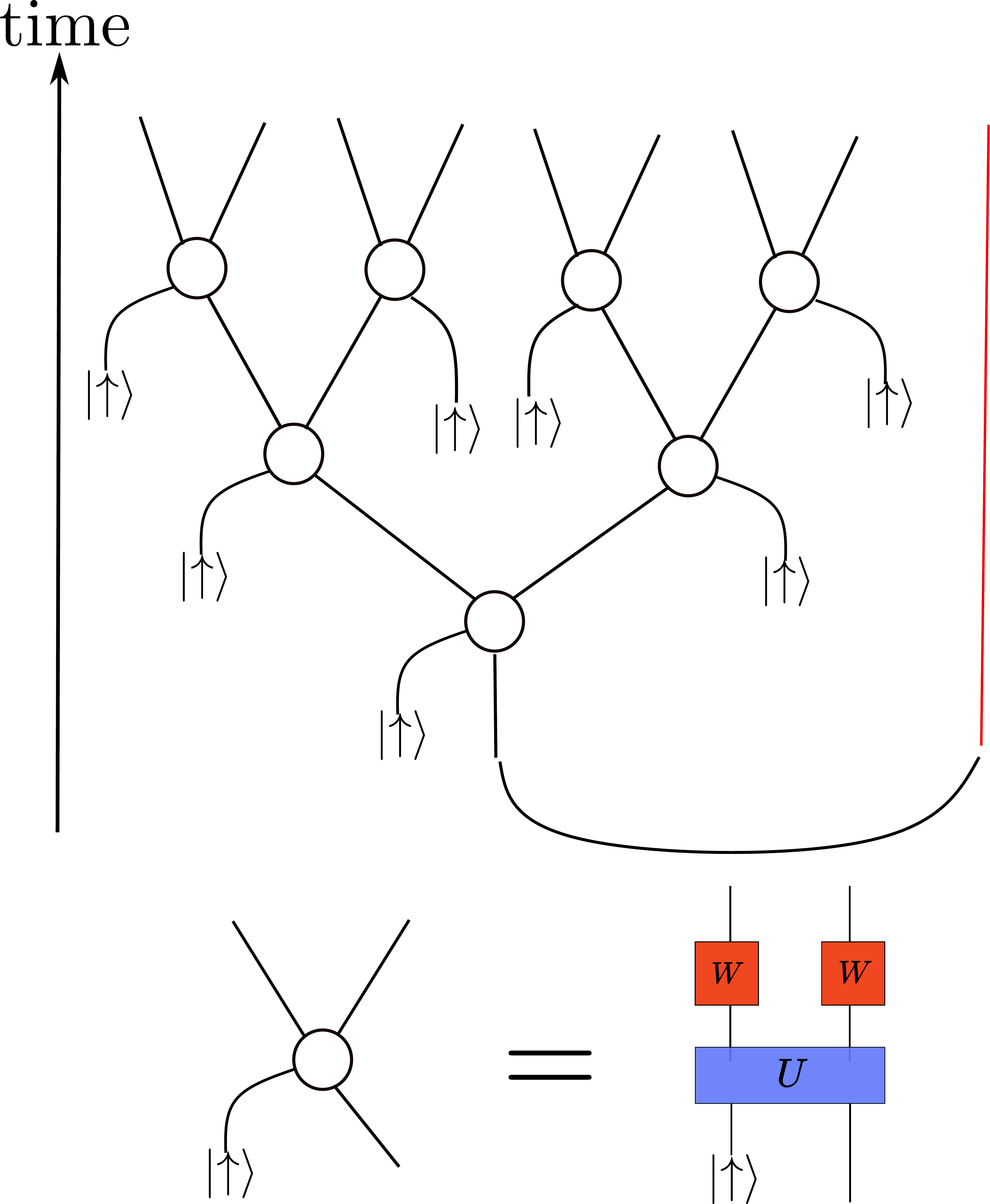}
    \caption{Schematic picture of the proposed expansion protocol. We use the $\cup$-shaped connection at the initial time to represent the initial GHZ state between one system spin and the reference spin. The world line of the reference spin is shown by the red line.  A possible choice for the design of each node, following Fig.~\ref{fig:tree_model}, is shown at the bottom.} 
    \label{fig:exp_cartoon}
\end{figure}

The critical properties of the expansion process with forced measurements have been solved in Sec.~\ref{sec:force}, since they map to those of the collapse process FMPT (the same ensemble of tensor networks arises in both cases, with opposite choices for the arrow of time). 
We do not solve the expansion process with true measurements here.
Unlike for forced measurements, reversing the arrow of time does not map the expansion process with true measurements to the collapse process with true measurements. 
Instead, reversing the arrow of time maps the expansion process with true (weak\footnote{Note that in the expansion process defined in Sec.~\ref{sec:model} there are only weak measurements, no strong measurements.}) measurements to a ``mixed'' version of the collapse process, in which the weak measurements are true measurements, but the projective measurements are forced.  This equivalence can be shown using the formulas in Sec.~\ref{sec:collapseTN}.

Thus, formally the expansion process with true measurements is an intermediate case between the two cases that were solved analytically in the previous sections. 
It would be interesting to understand how this intermediate case differs in its critical properties from the other two cases. 
A naive guess is that it has qualitatively similar scaling properties, with some value of the critical exponent $\lambda$. We leave an investigation of such mixed dynamics to the future. 

Instead, we comment  on the implementation of the measurement phase transition in the expansion process on a quantum device. 
Such an experiment does not require any postselection, in contrast to usual versions of the MPT. 
See Refs.~\cite{Gullans_Scalable_2020,Noel_Measurement_2022,Ippoliti_postselection_2021,Koh_Experimental_2022,Li_Cross_2022,garratt2022measurements}
for recent approaches to the statistical challenge associated with postselection.
In particular, Refs.~\cite{Li_Cross_2022,garratt2022measurements} discuss approaches involving ``hybrid observables'' that mix classical computations and quantum measurements. 
Our proposal below follows the spirit of these works, as it makes use of the efficient classical computability (given the measurement record) of certain expectation values. 
Here, the tree structure means that these classical computations are particularly simple.
It would also possible to study the collapse process experimentally: we comment on this in Sec.~\ref{sec:concl}.

The discussion in the remainder of this section is not restricted to any particular design of the ``node’’ interaction, so we will abstract away from the particular models we have considered so far, and imagine a more general version of the expansion process.
For example, it is not necessary that the interaction unitaries be random: 
we could consider a protocol in which the only randomness comes from the measurement outcomes.\footnote{It is also possible to replace weak measurements with projective ones, at the cost of involving more spins in the node. (This equivalence is a  consequence of the fact that a weak measurement can be effected using an additional ancilla spin that gets projectively measured.)}

Extrapolating from numerical explorations of the collapse process, it is likely that clear evidence of the phase transition could be obtained from an experiment on a relatively small number of qubits.

We consider an expansion process that starts with the single initial ``system'' spin in a maximally entangled (GHZ) state with a ``reference'' spin, denoted $R$ \cite{Gullans_Scalable_2020}.   (Performing a partial trace over the reference spin produces the maximally mixed initial state of the system spin, as we assumed when we introduced the expansion process in the Introduction.)
We run the expansion process for the system, recruiting  additional spins at each time step. 
 A cartoon of this dynamics is illustrated in Fig.~\ref{fig:exp_cartoon}, where the $\cup$-shaped connection at the bottom represents the initial entangled state of the ``system'' and ``reference'' spins.
We will detect the MPT using measurements of the reference $R$.

At the end of the expansion process we have an entangled state $\ket{\Psi}$ of  $R$ and the $2^t$ system spins.  
The Schmidt values for a bipartition of this state between $R$ and the system are 
$\{\sqrt{1-Z}, \sqrt{Z}\}$, 
where $Z$ is the order parameter that we have focused on in previous sections.
Tracing out the system, the reduced density matrix for the reference $\rho_R$ has eigenvalues ${1-Z}$ and $Z$.

To begin, imagine an idealized experiment with no uncontrolled noise. 
In a given run of the experiment we obtain a measurement record ${\mathbf{m}}_W$.
The final state $\rho_R$ of the reference will depend on ${\mathbf{m}}_W$.
However the recursive structure of the tree means that 
a very straightforward processing of the experimental data ${\mathbf{m}}_W$ allows us to deduce this final state $\rho_R$.
The computation simply iterates a recursion relation for $2\times 2$ density matrices. This recursion has exactly the same structure as 
Eq.~\ref{eq:nonlinearrecursionrho}: we start at the leaves of the tree and iterate towards the trunk (in the present setting this process corresponds to iterating backwards in time).\footnote{The starting state for this recursion corresponds to maximally mixed states for each of the ``leaves’’ of the tree.  For each node, the node tensor $t$ used for the recursive step depends on the measurement outcomes that were obtained at that node.}
Notably, the computational effort is proportional only to the number of nodes in the tree: 
it is sufficient to work with only $2\times 2$ density matrices, rather than a many-body quantum state.

Let us write $\rho_R$ in the form
\be
\rho_R= ( 1 + \vec n \cdot \vec \sigma)/2,
\ee
where $\vec{\sigma}$ is the vector of Pauli matrices, and 
\be
|\vec n| = 1 - 2Z.
\ee
In the idealised experiment,  $\rho_R$ can be computed analytically from the measurement record in each run, and one could in principle obtain observables like the order parameter $\langle Z\rangle$ directly from this computation. 
We will denote this estimate as $\langle Z\rangle_S$, since it makes use of measurement data for the system $S$ alone (not for the reference spin).
In reality, it is crucial to make a further measurement of the reference spin in each realization, in order to obtain independent confirmation of the result of the density matrix reconstruction. 
A conceptually simple way to do this is as follows. 

In each run one has  the recursively computed estimate of the polarization vector  $\vec n$ and therefore of its normalized version $\hat n$.
One can now make a Pauli measurement in the appropriately rotated frame, i.e.\ a measurement of  ${\hat n \cdot \vec \sigma}$.
This measurement yields an outcome $\tau=\pm 1$, 
with the outcome $\tau = -1$ having a probability $Z$.
Therefore these measurements give us an estimate of the average value of the order parameter $Z$:
\be
\langle Z \rangle_{SR} = \f{1}{2} \left(  1- \langle \tau \rangle \right),
\label{eq:Ztau}
\ee
where the averaging is over different runs of the experiment. The subscript on $\langle Z \rangle_{SR}$ indicates that this estimate   combines measurement data from both the system $S$ and the reference $R$.
$\langle Z \rangle_{SR}$ should agree with the estimate $\langle Z \rangle_{S}$ which uses measurement data from the system only.

Equation (\ref{eq:Ztau}) implies that the MPT for the expansion process corresponds to a phase transition in the \emph{predictability} of the final measurement on the reference spin. In the disentangling/strong measurement phase, the reduced density matrix $\rho_R$ for the reference spin is a pure state in the limit $k \rightarrow \infty$ of large tree depth. 
Consequently, the final measurement of the reference spin (the value of $\tau$) can be predicted with perfect accuracy by someone keeping track of the measurement record of the system spins (and using them to perform the recursive calculation of $\rho_R$). 
On the other hand, in the entangling/weak measurement phase the reference spin remains in a mixed state even as $k \rightarrow \infty$, and consequently the final measurement on the reference spin cannot be predicted with perfect accuracy.

The protocol described above requires one to make a spin measurement in an arbitrary basis defined by $\hat{n}$, which might be challenging.
However, even if measurements can only be made in a fixed basis, say $\sigma_z$, the average value of the order parameter, $\langle Z \rangle$, can still be extracted. Specifically,
\be
\langle Z \rangle = 
\f{1}{2} \left( 1 - \left\langle \f{ \tau }{ \hat n_z  } \right\rangle \right),
\ee
where $n_z$ is the $z$-component of the vector $\vec{n}$ that is calculated theoretically for each run,  ${\tau=\pm 1}$ is the $\sigma_z$ measurement outcome for the reference, and again the average is over runs. The value of $\langle Z \rangle$ calculated in this way using the experimental measurements of $\sigma_z$ should exhibit the critical scaling of the MPT.

\section{Outlook}
\label{sec:concl}

In this paper we have defined and explored the measurement-induced entanglement phase transition in  dynamical quantum trees.
We have adopted the perspective of purification  \cite{Gullans_Dynamical_2020}, so that the entanglement  transition is characterized by a single ``order parameter'' $Z_k$ that describes the purity of the final state after a $k$-step protocol of measurement and unitary evolution.

Our primary contribution is to extend the approach of Ref.~\cite{Nahum_Measurement_2021} in order to provide analytical solutions for a transition (for the collapse process)  not only for forced measurements, but also for real measurements.  
We have shown that a transition exists in both cases (the FMPT and the MPT), with distinct nontrivial values of the critical measurement strength $\theta_c$. 
The value of $\theta_c$ is somewhat smaller in the real measurement case, which in the language of purification  arises because real (weak) measurements are more purifying than forced ones.  
At the most basic level,  the MPT and FMPT exhibit similar critical behavior near the transition, as captured by Eqs.~\ref{eq:Ztyptheta_FMPT}--\ref{eq:Ztypk} and Eqs.~\ref{eq:ztypthetaMPT}--\ref{eq:ztypk_MPT},
but there are universal differences between the two types of transition.

The most striking difference between the MPT and FMPT is in the values of the critical exponent 
$\lambda$, 
which characterizes the breadth of the probability distribution for the order parameter $Z$ (or equivalently for the entanglement) at the critical point.
The MPT has the larger value ${\lambda = 1}$, as compared with ${\lambda = 1/2}$ for the FMPT.
This difference implies a narrower distribution of $Z_k$ values in the MPT case for a given tree size $k$.

The models are analysed using an auxiliary 
recursion relation for $Z$.
Interestingly, the value ${\lambda = 1}$ lies precisely at the boundary between two classes of phase transition
for such recursion relations.
The analysis here shows that the value
${\lambda=1}$ is protected for a larger class of models, in which the node operations have a statistical invariance under unitary rotations. 
But it is possible that modifications to the quantum tree model that take it outside this class of models could push $\lambda$ for the MPT to values larger than unity.
In such a model, the MPT and FMPT would show more drastically different universal properties.   (According to the arguments of Ref.~\cite{Nahum_Measurement_2021}, at ${\lambda > 1}$ the onset of the order parameter near the transition is as a power law, rather than exponential as in the model studied here.) 
Studying models in this broader class may therefore be a promising goal. 
These differences are also interesting in relation to possible field theories for the~MPT.

One could similarly ask whether smaller values of the exponent $\lambda$ are realizable when the statistical unitary invariance property of the current models is relaxed. This property enforces $\lambda=1$ for the  collapse process with real measurements and $\lambda=1/2$ for the  processes with forced measurements (at their respective critical points).
Heuristically, smaller $\lambda$ corresponds to stronger disorder and a broader distribution of entanglement.
Do the Haar-invariant tree ensembles correspond to the strongest possible disorder, or can other critical models realize broader order parameter distributions?

It is worth pointing out that our model contains no randomness in the time or location of measurements (in the FMPT case there is no randomness in the measurement outcomes either). Nonetheless, a phase transition still exists as a function of the measurement strength. 
Forthcoming work shows that a completely uniform tree tensor network, with no randomness of any kind, can also show an entanglement transition \cite{deterministictreeinprep}.
The model studied here, and extensions of it, may provide insight into the role of different kinds (and strengths) of randomness at the MPT.

An outstanding task that we have not tackled here is to develop a recursive approach to the expansion process with real measurements. 
Finding  an analytical solution to this problem would be interesting.
Does this process correspond to a phase transition of the same general type as those studied here, with some value of the critical exponent $\lambda$?

In Sec.~\ref{sec:expansion} we suggested a schematic experimental implementation of the MPT for the expansion process. 
It would also be possible to realize the collapse process experimentally. 
One general approach \cite{Li_Cross_2022} is to compare the evolution of different initial states that are conditioned on the same measurement outcomes.
In the collapse process, it is sufficient  to compare the output single-spin state for distinct initial many-body  states.
In outline, one initial state 
is run experimentally and the other initial state  is simulated classically, enforcing the same outcomes. 
The average distance between the outputs can then be estimated with an appropriate measurement, and can serve as an  order parameter (in the strong monitoring phase, the tree tensor network has a single dominant singular value, and so projects almost all inputs onto the same output).
As in the case of Clifford circuits \cite{Li_Cross_2022}, the initial state that is run experimentally can be a state that is not efficiently representable classically.

Separately, it would also be interesting to understand whether the expansion or collapse processes are related to any natural quantum information task. 
One could also explore 
quantum versions of classical message-broadcasting processes with a tree-like structure, with potential transmission errors on the bonds of the tree. These classical processes can show nontrivial phase transitions
in the information communicated from the root to the leaves
\cite{mezard2006reconstruction}. 

The results here were obtained using a recursion relation for a random variable $Z_k$ characterizing the entanglement (or purity) of a given tree. 
An open question is how to recover them using the formalism of the effective ``lattice magnet'' that can be obtained using the replica trick by integrating out all the random unitaries~\cite{Vasseur_Entanglement_2019,Jian_Measurement-induced_2020,Bao_Theory_2020,Zhou_Emergent_2019,Lopez-Piqueres_Mean-field_2020,Nahum_Measurement_2021}.

 Finally, we note that we have only studied the entanglement between the root and leaves of the tree, but it would also be interesting to analyze the entanglement between subsets of the leaves. For example, one could consider the final state produced by an expansion process for which the initial state is pure. In this case one can think of the final state of the leaves as a 1D wavefunction analogous to a multi-scale entanglement renormalization ansatz (MERA) state. For such a state the entanglement entropy between subsets of the leaves also reflects the transition. In the weak monitoring phase, we expect a logarithmic scaling of the entanglement entropy with subsystem size \cite{pfeifer2009entanglement, Nahum_Measurement_2021}.

\acknowledgments 
The authors are grateful to Bernard Derrida, Sthitadhi Roy, and Jonathan Ruhman for helpful discussions. 
Part of this work was performed at the Aspen Center for Physics, which is supported by National Science Foundation grant PHY-1607611.

\appendix

\section{Gauge transformation of tree}
\label{app:gaugetransf}

In Sec.~\ref{sec:collapse} we stated that it did not matter whether the  measurements were performed in the $S_z$ basis, or in randomized bases. In this Appendix we show this using the expressions in Sec.~\ref{sec:collapseTN} for the probability of a given tree. We consider the case of true measurements (the forced measurement case is even simpler) for the collapse process. The same logic applies to the expansion process.

Abusing notation, in this Appendix (only) we will denote a given tree by
\be
T = T(U,  u, v, \sigma).
\ee
Here $U$ denotes not a single unitary but the complete set of two-spin unitaries appearing in the tree,
$u$ denotes the set of single-spin unitaries appearing in the Kraus operators (which fix the measurement bases for the weak measurements) and  $v$ denotes a set of single-spin unitaries that fix the bases for the strong measurements.  Finally $\sigma$ is the set of measurement outcomes. We will also use $U$, $u$ etc. for individual unitaries --- this should be clear from context.

First, note that, because of the structure of the complete tensor network for $T$, it is possible to absorb almost all of the $u$s and $v$s into a redefinition of the $U$s.  
For example, in  the expression ${K=u K_\text{diag} u^\dag}$ for the Kraus operator on a bond, we can absorb the $u$ into the $U$ above the bond, and the $u^\dag$ in to the $U$ below.
Similarly each $v$   can be absorbed into the $U$ from the same node.
In this process, a given two-site unitary $U_r$ (here $r$ denotes a node) transforms as
\be
U_r \longrightarrow V_r(u,v) U_r W_r(u,v).
\ee
Where $V_r$ and $W_r$ are some unitaries that depend on the adjacent $u$s and the adjacent $v$. (The details do not matter.)
The only $u^\dag$s that do not get absorbed in this process are the ones at the  initial time  from the lowest layer of $K$s. If we denote the product of these by $X(u)$, then the above ``gauge transformation'' gives us the simple identity
\be\label{eq:gaugetransformation}
T( U, u, v, \sigma) =  T\lf V(u,v) U W(u,v), \mathbf{1}, \mathbf{1}, \sigma\ri  \times X(u).
\ee
Again we use a schematic notation where site indices are dropped.

The final factor, $X(u)$, will drop out because (i)  we take the initial state to be maximally mixed, and (ii)  we consider observables $F(T)$ that are invariant under a unitary transformation at the base of the tensor network. That is, regarding $T$ as a $2\times M$ rectangular matrix (see Sec.~\ref{sec:collapseTN}),
\be\label{eq:Finvariance}
F(T) = F (TY)
\ee
for any  $M\times M$ unitary $Y$. For example, any function of the order parameter $Z$ has this property, since the singular values of $T$ are unchanged by a unitary rotation.

Now we wish to show that the averages of $F$ are the same in two different ensembles: 
first, the ensemble where the single-site unitaries $u$ and $v$ are Haar-random, and second, the ensemble where they are all fixed to the identity. The latter corresponds to taking all measurements in the $S_z$ basis. 
(The case where the $u$s are random while the $v$s are fixed, or vice versa,  can also be shown to be equivalent, in a similar way).

First  consider the case where the bases are randomized. Then expectation values have the schematic form
\be
\left\langle F\right\rangle=\left\langle\left\langle \left\langle F\lf T( U,  u, v, \sigma) \ri  \right\rangle_\sigma \right\rangle_U \right\rangle_{u,v}
\ee
The unitary averages involve the Haar measure. 
The $\sigma$ average has to be the innermost one, and it is done with the Born rule probability, which is a function of $T$. 
We write this schematically as a function of $T$ (see  Sec.~\ref{sec:collapseTN}):
\be
p(\sigma|U,u,v) = H(T(U, u, v, \sigma)).
\ee
Since we start with a maximally mixed state, the function $H$ is expressed in terms of the trace of $TT^\dag$, so it obeys property (\ref{eq:Finvariance}). Therefore, using (\ref{eq:gaugetransformation}),
\begin{align}
p(\sigma|U,u,v) & = H(T(V(u,v)UW(u,v), \mathbf{1}, \mathbf{1} , \sigma))\\
& = p(\sigma| V(u,v)UW(u,v), \mathbf{1},\mathbf{1}).
\end{align}
Now return to the average. Let us label the inner average by the probability function used for the average over $\sigma$:
\begin{align}
\left\langle F\right\rangle & =\left\langle\left\langle \left\langle F\lf T( U,  u,v, \sigma) \ri  \right\rangle_{\sigma;\, p(\sigma|U,u,v)} \right\rangle_U \right\rangle_{u,v}.
\end{align}
Inside the average, we use the identity (\ref{eq:gaugetransformation}).
\begin{align}
\left\langle F\right\rangle & =\left\langle\left\langle \left\langle F\lf \widetilde T \ri  \right\rangle_{\sigma; \, p(\sigma|U,u,v)} \right\rangle_U \right\rangle_{u,v},
\end{align}
where $\widetilde T$ stands for ${\widetilde T= T( V(u,v)UW(u,v),  \mathbf{1},\mathbf{1}, \sigma)}$.
We also use the identity for $p$.
\begin{align}\notag
\left\langle F\right\rangle & =\left\langle\left\langle \left\langle F\lf \widetilde T \ri 
 \right\rangle_{\sigma; \, p(\sigma|V(u,v)UW(u,v), \mathbf{1},\mathbf{1})} \right\rangle_U \right\rangle_{u,v}.
\end{align}
Now we define $U'=V(u,v)UW(u,v)$.
We note that for any fixed  $u$ and $v$, the distribution of   $U'$ is independent of $u$ and $v$ and is Haar.
\begin{align}
\left\langle F\right\rangle & =\left\langle\left\langle \left\langle F\lf T( U',  \mathbf{1},\mathbf{1}, \sigma) \ri  \right\rangle_{\sigma; \, p(\sigma|U', \mathbf{1},\mathbf{1})} \right\rangle_{U'} \right\rangle_{u,v}.
\end{align}
Nothing inside the outer average depends on $u$, $v$ so we can drop the average over these quantities:
\begin{align}
\left\langle F\right\rangle & =\left\langle \left\langle F\lf T( U',  \mathbf{1},\mathbf{1}, \sigma) \ri  \right\rangle_{\sigma; \, p(\sigma|U', \mathbf{1},\mathbf{1})} \right\rangle_{U'} 
\end{align}
This is the average in the ensemble where all the measurements are in the $S_z$ basis.
This establishes the claim.

Let us also confirm a similar invariance which we invoked at the beginning of Sec.~\ref{sec:real_linear}. Consider a single node of the collapse process, with real measurements, with a definite initial state of the form ${\rho_1\otimes \rho_{2}}$.
We check that the probability distribution for the output density matrix,
\be
\rho \propto t [\rho_1\otimes\rho_{2}] t^\dag,
\ee
is left unchanged by a basis rotation for one of the initial spins, e.g.
\be
\rho_1 \rightarrow w \rho_1 w^\dag.
\ee
This invariance can be seen by manipulations similar to the above. The node tensor $t$, defined in Sec.~\ref{sec:collapseTN},
depends on a two-site unitary $U$, on single-site unitaries $u^1$, $u^2$ appearing in the Krauss operators, and on measurement outcomes ${s=(\sigma_1,\sigma_2, \sigma_2')}$. Schematically, $\rho$ is a function of all these and the initial states. Let us write
\be
\rho = \rho(s,X), \text{ where } X = (U, u^1, u^2, \rho_1, \rho_2).
\ee
Let us denote the probability of outcomes $s$, given the initial states and the unitaries, as $p(s|X)$ (see Eq.~\ref{eq:real_outcomep} for the expression).
Let us also define
\begin{align}
X_w & =  (U, u^1, u^2, w\rho_1 w^\dag, \rho_2), \\
\widetilde X_w & = (U[w\otimes \mathds{1}], w^\dag u^1, u^2, \rho_1 , \rho_2).
\label{eq:tildeXw}
\end{align}
Then the structure of the tensor network for $\rho$ gives the following invariances,
\begin{align}
\rho(s,X_w) & =  \rho(s,\widetilde X_w),
&
p(s|X_w) & =  p(s|\widetilde X_w).
\end{align}
As a result, the probability distribution of $\rho$, for given input states, is unchanged by the rotation $\rho_1\rightarrow w \rho_1 w^\dag$. To see this, consider an arbitrary expectation value involving $\rho$, with the initial state $(w\rho_1 w^\dag)\otimes \rho_2$:
\begin{align}
\langle F(\rho) \rangle
& =
\langle\langle F(\rho(X_w)) \rangle_{s;p(s|X_w)}  \rangle_{U,u^1,u^2} 
\\ & = 
\langle\langle F(\rho(\widetilde X_w)) \rangle_{s;p(s|\widetilde X_w)}  \rangle_{U,u^1,u^2}.
\end{align}
Looking at Eq.~\ref{eq:tildeXw} we see that $w$ and $w^\dag$ can be absorbed into the Haar-random unitaries $U$ and $u^1$ without any change to their distribution, so that $F(\rho)$ is independent of $w$. 
That is, the probability distribution of the output $\rho$ depends only on the eigenvalues of $\rho_1$ and not on its eigenvectors (and similarly for $\rho_2$).

\section{$\lambda^{\ast}$ parameter: real measurement case}
\label{appendix:lambda}

In this appendix we derive an identity, used in 
Sec.~\ref{sec:real_linear}, which fixes the value of  $\lambda^{\ast}$ for the true measurement case. We start with Eq.~\ref{real_identity},
\be
\sum_{s}\langle p(s)(A^{\lambda^{\ast}}_1(s)\ln A_1(s)+A^{\lambda^{\ast}}_2(s)\ln A_2(s))\rangle =0, \label{eqn:A1}
\ee
where 
$s=(\sigma_1,\sigma_2,\sigma_2')$ labels measurement outcomes, and
we recall from Sec.~\ref{sec:real_linear} and Sec.~\ref{sec:collapseTN} that
\begin{align}
A_1(s) &= \frac{\left|t_{11}^{1}t_{21}^{2}-t_{21}^{1}t_{11}^{2}\right|^2}{\left(|t_{11}^{1}|^2+|t_{11}^{2}|^2\right)^2},&
    A_2(s) &=\frac{\left|t_{11}^{1}t_{12}^{2}-t_{12}^{1}t_{11}^{2}\right|^2}{\left(|t_{11}^{1}|^2+|t_{11}^{2}|^2\right)^2},
    \label{eq:Aappendix}
    \end{align}
and
\be\label{eq:psappendix}
p(s) = |t^1_{11}|^2 + |t^2_{11}|^2,
\ee
and finally
\be\label{eq:tappendix}
 t^a_{cd} =
  t(\sigma_1,\sigma_2,\sigma_2')^a_{cd} =
 \left[U \, (K^{(1)}_{\sigma_1}\otimes K^{(2)}_{\sigma_2})\right]^{a\sigma_2'}_{cd}.
\ee
Since the average over measurements has been separated out and written explicitly in terms of ${p(s)=p(\sigma_1,\sigma_2, \sigma_2')}$, 
the angle brackets in Eq.~\ref{eqn:A1} represent only the averages over the Haar-random unitary $U$ and  the random bases in the $K$ operators.\footnote{We could write these operators as e.g. 
\be
K^{(1)}_\sigma = u^{(1)}
K^{z}_\sigma u^{(1)\dag},
\ee
where $K^{z}$ is non-random and diagonal in the $z$ basis, and $u^{(1)}$ is a Haar random $2\times 2$ unitary.}
Therefore, \textit{these} averages are taken with a distribution that is invariant under unitary transformations on any index of the tensor $t$.

Eq.~\ref{eqn:A1} means that the quantity we need to evaluate is not $\langle A_i^{\lambda}\rangle$ alone, as the forced measurement case, but $\langle p(s) A_i^{\lambda}(s)\rangle$. 
Since $A_1$ and $A_2$ are statistically equivalent, it suffices to examine the average involving $A_1$.
Using Eqs.~{\ref{eq:Aappendix}--\ref{eq:tappendix}}, 
\begin{align}
  \left\langle p(s) A_1^{\lambda}(s)\right\rangle &= \left\langle \left(|t_{11}^{1}|^2+|t_{11}^{2}|^2\right) \frac{\left|t_{11}^{1}t_{21}^{2}-t_{21}^{1}t_{11}^{2}\right|^{2\lambda}}{\left(|t_{11}^{1}|^2+|t_{11}^{2}|^2\right)^{2\lambda}}\right\rangle
  \nonumber \\
  &=\left\langle \frac{\left|t_{11}^{1}t_{21}^{2}-t_{21}^{1}t_{11}^{2}\right|^{2\lambda}}{\left(|t_{11}^{1}|^2+|t_{11}^{2}|^2\right)^{2\lambda-1}}\right\rangle. \label{eqn:A2}
\end{align}
This can be analyzed by a similar method to Ref.~\cite{Nahum_Measurement_2021}. Let $M$ be the matrix ${M_{ab} \equiv t^a_{b1}}$, in terms of which 
\begin{align}
  \left\langle p(s) A_1^{\lambda}(s)\right\rangle 
  &=\left\langle \frac{\left| \det M \right|^{2\lambda}}{\left(|M_{11}|^2+|M_{21}|^2\right)^{2\lambda-1}}\right\rangle. \label{eqn:Mappendix}
\end{align}
We consider the  singular value decomposition of $M$,
\be
M_{ab} =
 \sum_{\mu =1,2} w_{a\mu}\eta_{\mu}v_{\mu b}, \label{eqn:A3}
\ee
where $w$ and $v$ are unitary, and, with probability 1, both singular values are greater than zero.
Then Eq.~\ref{eqn:A2} becomes
\be
  \left\langle p(s) A_1^{\lambda}(s)\right\rangle = \left\langle\left\langle\frac{\eta_1^{2\lambda}\eta_2^{2\lambda}}{\left(\eta_1^2\left|v_{11}\right|^2+\eta_2^2\left|v_{21}\right|^2\right)^{2\lambda-1}}\right\rangle_{v}\right\rangle_{\eta}. \label{eqn:A4}
\ee
Here the averages are over the singular values $\eta$ and the unitary matrix $v$.  
Because of the invariance property of the $t$ distribution mentioned above, the $v$ average is simply a Haar average, which reduces to a uniform average of the complex  vector $v_{a1}$ over the unit sphere, as in  \cite{Nahum_Measurement_2021}. Performing the sphere average (see  Eqs.~\ref{eq:Cmudefn},~\ref{eq:Cmudefn} below for the necessary formula) gives
\be
\left\langle p(s) A_1^{\lambda}(s)\right\rangle=
\left\langle\frac{\eta_1^{4-2\lambda}\eta_2^{2\lambda}-\eta_1^{2\lambda}\eta_2^{4-2\lambda}}{(2-2\lambda)(\eta_1^2-\eta_2^2)}\right\rangle_{\eta}. \label{eqn:A5}
\ee
This expression is sufficient to obtain the desired identity, even without performing the $\eta$ average explicitly. Differentiating with respect to $\lambda$ and taking the limit $\lambda \to 1$, we obtain
\be
\langle p(s) A_1\ln A_1\rangle =0, \label{eqn:A9}
\ee
for any outcome. Thus we prove Eq.~\ref{real_identity}. And the critical point is given by Eq.~\ref{eq:vzeroMPT}.

\section{Exact calculation of averages}
\label{app:averages}

The calculations in the previous appendix required only the unitary invariance property of the node averages, and did not use the specific details of the $t$ tensor. 
But for our model it is possible to calculate these averages exactly. We will evaluate the average
\be\label{eq:Xkldefn}
X_{\kappa, \lambda}(\theta) \equiv  \left\langle 
p(s)^\kappa A_1(s, \theta)^\lambda
\right\rangle.
\ee 
As above, the average is over the two-site unitary $U$ and the Kraus operators in the node tensor $t$. The result is independent of the measurement outcomes $s=(\sigma_1,\sigma_2,\sigma_2')$, so we can take ${s=(\uparrow,\uparrow,\uparrow)}$ without loss of generality. The analogous expression with $A_2$ in place of $A_1$ is also equal to $X_{\kappa,\lambda}$. 

Before examining Eq.~\ref{eq:Xkldefn}, we define a basic average that we will need repeatedly,
\be\label{eq:Cmudefn}
C_\mu(\alpha,\beta)\equiv 
\left\langle
\f{1}{(\alpha |z_1|^2 + \beta |z_2|^2)^\mu} 
\right\rangle_{\vec z}.
\ee
Here ${\vec z=(z_1,z_2)}$ is a complex  unit vector that is averaged uniformly over the sphere. The average can be performed by standard means (see e.g. \cite{Nahum_Measurement_2021}) giving
\be\label{eq:Cmuresult}
C_\mu(\alpha,\beta)= \f{\alpha^{1-\mu}-\beta^{1-\mu}}{(1-\mu) (\alpha-\beta)}.
\ee

To get started we again write $X_{\kappa,\lambda}$ in terms of the matrix 
\be
M_{ab} = t^a_{b1}
\ee
(see Eq.~\ref{eq:tappendix} for the definition of $t$) as
\be
\label{eq:Xkl}
X_{\kappa, \lambda} =
 \left \langle 
\f{ |\det M|^{2\lambda}  }
{ \lf |M_{11}|^2 + |M_{21}|^2\ri^{2\lambda-\kappa} }
\right\rangle.
\ee
Let us look at the structure of $M$. 
Suppressing the ``$\uparrow$'' subscripts on the $K$s that indicate the measurement outcomes, this matrix has the form
\begin{align}
M_{ab} 
=
U^{a 1}_{b' c'} K^{(1)}_{b' b} K^{(2)}_{c'1}.
\end{align}
The final factor,   $y_{c'}\equiv K^{(2)}_{c'1}$, has a single index, i.e. it is a random vector:
\be\label{eq:yvector}
y_{c'} = K^{(2)}_{c'1} = u_{c'd} 
\left(
\begin{array}{cc}
 \sin\theta &   0   \\
  0   & \cos\theta   
\end{array}
\right)_{dd'}
u^\dag_{d'1}.
\ee
Because $U$ is anyway Haar-random (and independent of $K^{(2)}$),
the orientation of this vector does not matter.
We are free to replace $y_{c'}$ with a vector that points in the ${(1,0)}$ direction, without changing the distribution of $M$. 
However, we must keep track of the norm of $y$. 
Let us call this norm $R$. 
Since the leftmost unitary on the RHS of (\ref{eq:yvector}) does not change the norm,
\be
R = \left|  
\left(
\begin{array}{cc}
 \sin\theta &   0   \\
  0   & \cos\theta   
\end{array}
\right)  \vec z \, \right|.
\ee
Here, $\vec z$ is a \textit{uniformly random} vector (previously written as $u^\dag_{d'1}$). Note that 
\be
R^2 = \sin^2\theta \, |z_1|^2 + \cos^2\theta \, |z_2|^2
\ee
is a random variable of the type appearing in the denominator in Eq.~\ref{eq:Cmudefn}.

After the simplification above, we have 
\be\label{eq:Mfirstsimplification}
M_{ab} = R \times U^{a1}_{b'1} K^{(1)}_{b'b},
\ee
where all three elements are statistically independent. 

Next let us make a singular value decomposition of $U$, or rather of the sub-block of $U$ that appears in $M$:
\begin{align}\label{eq:Ublocksvd}
U^{a1}_{b'1} & =  (v H w)_{ab'},
&
H & = 
\left(
\begin{array}{cc}
\eta^U_1  & 0    \\
0  & \eta^U_2  
\end{array}
\right).
\end{align}
Here $v$ and $w$ are Haar-random $2\times 2$ unitary matrices.
(We have denoted the singular values by $\eta^U_\mu$ to distinguish them from the singular values of $M$ itself, which were denoted $\eta_\mu$ in Sec.~\ref{appendix:lambda}.)
The distribution of the singular values $\eta^U_\mu$ is provided in Ref.~\cite{kieburg2016singular}. Defining $s_i = (\eta^U_i)^2$, we have ${0\leq s_i\leq 1}$ and 
\be\label{eq:sdistr}
P(s_1, s_2)\,ds_1\, ds_2 = 6(s_1-s_2)^2 \,ds_1 \, ds_2. 
\ee
The remaining factor in the definition of $M_{ab}$ involves the matrix $K^{(1)}$, which is:
\be\label{eq:K1expr}
K^{(1)} = u' \left(
\begin{array}{cc}
 \sin\theta &   0   \\
  0   & \cos\theta   
\end{array}
\right) 
u'^\dag,
\ee
where $u'$ is yet another $2\times 2$ Haar-random unitary matrix.  Altogether, from (\ref{eq:Mfirstsimplification}), (\ref{eq:Ublocksvd}) and (\ref{eq:K1expr}) we have
\be\label{eq:Massemble}
M = R \times v \, H \,w \, u' 
\left(
\begin{array}{cc}
 \sin\theta &   0   \\
  0   & \cos\theta   
\end{array}
\right) 
u'^\dag.
\ee

Now consider the numerator and denominator in (\ref{eq:Xkl}).
Since $\det AB = \det A \det B$, and the unitary matrices have unimodular determinant,
we have for the determinant of $M$
\be\label{eq:detsimplified}
|\det M| = \f{1}{2} R^2  \eta^U_1 \eta^U_2  \sin 2\theta.
\ee
From Eq.~\ref{eq:Massemble}
we see that the expression ${|M_{11}|^2 + |M_{21}|^2}$ is the 
 norm-squared of the vector
\be\label{eq:vectortobenormed}
R \times \left[  H \tilde w
\left(
\begin{array}{cc}
 \sin\theta &   0   \\
  0   & \cos\theta   
\end{array}
\right) 
{\vec z\phantom{.}'} \right],
\ee
where 
we have defined ${z'_a\equiv u'^\dag_{a1}}$, and
 the product $w u'$ has been written as $\tilde w$ ---  which is itself Haar-random --- 
 and finally the unitary rotation $v$ in (\ref{eq:Massemble}) has been dropped since it does not change the norm.
 \be\label{eq:MsqplusMsq}
 |M_{11}|^2 + |M_{21}|^2 = R^2 \left| 
 H \tilde w
\left(
\begin{array}{cc}
 \sin\theta &   0   \\
  0   & \cos\theta   
\end{array}
\right) 
{\vec z\phantom{.}'} 
 \right|^2.
 \ee 
 The unit vector ${\vec z\phantom{.}'}$ is again uniformly random.

The vector whose norm is taken in (\ref{eq:MsqplusMsq}) is obtained by applying the diagonal matrix $H$ of singular values to the complex vector $\vec x$,
\be\label{eq:xvectordefn}
\vec x \equiv \tilde w 
\left(
\begin{array}{cc}
 \sin\theta &   0   \\
  0   & \cos\theta   
\end{array}
\right) 
{\vec z\phantom{.}'}.
\ee
This vector can be written
\be\label{eq:xhatx}
\vec x = R' \hat x.
\ee
Here, the norm $R'$ is given by
\be
R'^2 = \sin^2\theta \, |z'_1|^2 + \cos^2\theta\, |z'_2|^2,
\ee
which is another random variable distributed in precisely the same way as $R^2$ above (meaning that  we will be able to use Eq.~\ref{eq:Cmuresult}).
The unit vector $\hat x$ in (\ref{eq:xhatx}) is uniformly distributed on the sphere, because of the Haar-random unitary $\tilde w$ in (\ref{eq:xvectordefn}). 

We have
\be
 |M_{11}|^2 + |M_{21}|^2 = R^2  R'^2 \left| H \hat x \right|^2,
\ee
giving 
\be\label{eqMMnorms}
 |M_{11}|^2 + |M_{21}|^2 =
 R^2  R'^2 \lf s_1 |\hat x_1|^2 + s_2 |\hat x_2|^2 \ri.
\ee
All three factors on the right hand side of (\ref{eqMMnorms}) are random variables of precisely the kind appearing in Eq.~\ref{eq:Cmudefn}, with $(a,b)=(\sin^2\theta, \cos^2\theta)$ for the first two factors, and $(a,b)=(s_1,s_2)$ for the third factor.

Combining Eq.~\ref{eq:detsimplified} and Eq.~\ref{eqMMnorms},
\begin{align}\notag
X_{\kappa,\lambda}  = &  
\lf \f{\sin 2\theta}{2} \ri^{2\lambda} 
\left\langle
R^{2\kappa} 
 \right\rangle_{\vec z}
 \left\langle
R'^{2\kappa-4\lambda} 
 \right\rangle_{\vec z'}  \\ & \quad
 \times
\left\langle
\left\langle
\f{(s_1s_2)^\lambda  }{\lf s_1 |\hat x_1|^2 + s_2 |\hat x_2|^2 \ri^{2\lambda-\kappa}}
\right\rangle_{\vec x}
\right\rangle_{s_1,s_2}
\end{align}
Using the definition (\ref{eq:Cmudefn}),
\begin{align}\notag
X_{\kappa,\lambda}  & = 
\lf \f{\sin 2\theta}{2} \ri^{2\lambda} 
C_{-\kappa}(s^2\theta,c^2\theta)
C_{2\lambda-\kappa}(s^2\theta,c^2\theta) \\
& \quad \quad\quad \quad\quad \quad \times
\left\langle
(s_1s_2)^\lambda
C_{2\lambda-\kappa}(s_1,s_2)
\right\rangle_{s_1,s_2},
\end{align}
where $(s^2 \theta,c^2\theta) = (\sin^2\theta, \cos^2\theta)$.
Finally we must integrate over $s_1$ and $s_2$ using (\ref{eq:sdistr}),
\begin{align}\notag
&\left\langle
 (s_1s_2)^\lambda
C_{2\lambda-\kappa}(s_1,s_2)
\right\rangle_{s_1,s_2}  = 
\\
& \quad\qquad\quad
\f{12}
{(1+\lambda)(2+\lambda)(2+\kappa-\lambda)(3+\kappa-\lambda)}.
\end{align}
Altogether this gives
\begin{align} \notag
& X_{\kappa,\lambda} = \\ \notag
&\f{3\times 4^{1-\lambda}
(\sin 2\theta)^{2\lambda}
(\cos2\theta)^{-2}
}{
(1+\kappa)(1+\lambda)(2+\lambda)(1+\kappa-2\lambda)
(2+\kappa-\lambda) (3+\kappa-\lambda)
}
\\ \notag
& \times
\lf  (\sin\theta)^{2+2\kappa} - (\cos\theta)^{2+2\kappa}   \ri
\\ & \times 
\lf  (\sin\theta)^{2+2\kappa-4\lambda} - (\cos\theta)^{2+2\kappa-4\lambda}   \ri.
\end{align}
The two cases of interest to us here are $X_{0,\lambda}$ (for the FMPT)  and $X_{1,\lambda}$ (for the MPT).
\begin{align} \notag
& X_{0,\lambda} = \\
& \f{-3\times 4^{1-\lambda}
(\sin 2\theta)^{2\lambda}
(\cos2\theta)^{-1}
\lf  (\sin\theta)^{2-4\lambda} - (\cos\theta)^{2-4\lambda}   \ri
}{
(1+\lambda)(2+\lambda)(1-2\lambda)
(2-\lambda) (3-\lambda)
}
\end{align}
and
\begin{align} \notag
& X_{1,\lambda} = \\
& \f{
-3 \times 2^{1-2\lambda} (\sin 2\theta)^{2\lambda} (\cos 2\theta)^{-1} 
\lf 
(\sin \theta)^{4-4\lambda} - (\cos \theta)^{4-4\lambda} 
\ri
}{
(1+\lambda)(2+\lambda)(3-\lambda)(4-\lambda)(2-2\lambda)
}
\end{align}
Some simple checks of these formulas are the values 
\begin{align}
X_{0,0} & = 1, &
X_{1,0} & = 1/8, &
X_{0,1} & = 1.
\end{align}
The second of these is the inverse of the number of possible measurement outcomes and follows from the normalisation ${\sum_s p(s)=1}$, and the third is a general result from the unitary invariance property.

The values that we need to locate the critical points are $X_{0,1/2}$ for the FMPT and $X_{1,1}$ for the MPT. Defining 
\be
\gamma \equiv \tan\theta,
\ee
these values are 
\begin{align}
X_{0,1/2} & = \f{128}{75} \f{\ln \gamma}{\gamma-1/\gamma}, 
&
X_{1,1} &=\f{1}{3} \f{\ln \gamma}{\gamma^2-1/\gamma^2}. 
\end{align}
For the FMPT we have the critical point equation
\be
X_{0,1/2}  = \f{1}{2}
\ee
which gives
\be
\theta_c^\text{FMPT} \simeq 1.42010054727632.
\ee
For the MPT the critical point equation is 
\be
X_{1,1} = \f{1}{16} 
\ee
which gives
\be
\theta_c^\text{MPT} \simeq 1.10010302468401.
\ee

\section{Finite pool-size effect}
\label{appendix:poolsize}

In this work we use the pool method to approximate the recursive evolution of the probability distribution of the density matrix in the quantum tree problem. 
For any fixed $k$, 
averages obtained by the pool method  should 
converge slowly to the true values as the pool size $N_\text{pool}$ tends to infinity. 
Therefore an important question is whether the pool size is large enough.

\begin{figure}[t]
    \centering
    \includegraphics[width = 1.0\columnwidth]{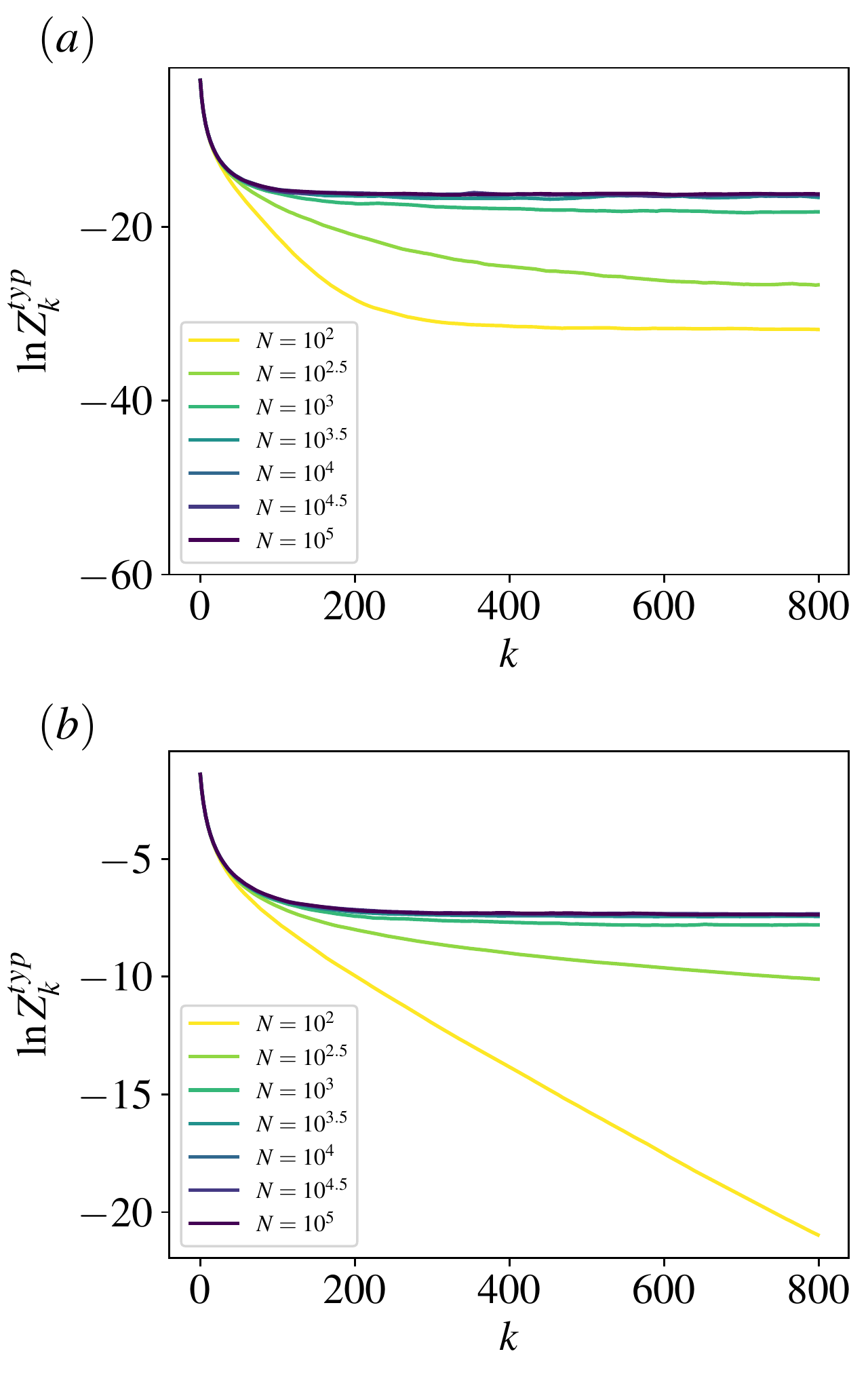}
    \caption{Convergence of results with increase of pool size. (a) The forced measurement case with $\ln(\theta_c-\theta) = -3.5$.  (b) The real measurement case with $\ln(\theta_c-\theta) = -3.9$. Here the relative error between the two  curves for the largest pool sizes shown in the figure, $10^{4.5}$ and $10^5$, is smaller than $1\%$. All error bars are smaller than the marker sizes.  
    }
    \label{fig:poolsize}
\end{figure}

To check whether the pool size in our study ($10^6$) is large enough, we choose $\theta$ close to the critical point and study the curves of $\ln Z^{typ}_k$ with different pool sizes for both the forced and real measurement cases. The result is shown in Fig.~\ref{fig:poolsize}. The relative error between the two  curves for the largest pool sizes shown in the figure, $10^{4.5}$ and $10^5$, is smaller than $1\%$. This convinces us that the pool size of $10^6$ used in the main text is sufficient.

\section{Order parameter scaling}
\label{appendix:infinite_z}

In this Appendix we discuss how to use the recursion relation for $G_k$ to extract the scaling of the order parameter in the entangling phase (but close to the transition). 
We focus on the case of true measurements, commenting on the differences for the simpler case of forced measurements.
The forced measurement case falls into the class of tree tensor networks for which order parameter scaling was discussed in Ref.~\cite{Nahum_Measurement_2021} using an analogy with an FKPP-like equation.

In the main text (Eq.~\ref{eq:recursionGGG}) we discussed the recursion relation for the generating function $G_k$ that follows from the linearized recursion relation for $Z$. 
We now give the nonlinear form of the recursion that allows for (1) higher-order terms in $Z_{k,1}$, $Z_{k,2}$ in Eq.~\ref{real_lin_recursion} and 
(2) a dependence of the outcome probability $p(s)$ on $Z_{k,1}$ and $Z_{k,2}$.

In this Appendix (only), we will use $\mathbb{U}$ to denote the collection of random unitaries involved in a given node.
Below, various  quantities ($A_i$, $F$, $p$) depend both on $\mathbb{U}$ and on $s$, but in order to shorten the formulas we will often suppress these arguments. 
However we will make explicit all dependency on the quantities $Z_1\equiv Z_{k,1}$ and $Z_2\equiv Z_{k,2}$.

We begin with the nonlinear version of Eq.~\ref{real_lin_recursion}:
\be
Z_{k+1} = A_i Z_{i} + F(Z_{1},  Z_{2}),
\ee
where the expansion of $F$ starts at order $Z^2$.
The sum over $i=1,2$ is assumed in the first term.
$A_i$ and $F$ depend on the set $s$ of measurement outcomes at the node, and the probability distribution $p_s$ for $s$ also depends on $Z$. To emphasize the $Z$ dependence of $p_s$ we will write it as:
\be
p_s(Z_1,Z_2).
\ee
The generating function of interest is 
\be
G_{k+1}(x) = 
\left\langle
\sum_s 
p_s(Z_1,Z_2) 
e^{
- e^{-x} \lf
A_i Z_{i} + F(Z_1,Z_2)
\ri
}
\right\rangle_{\mathbb{U},Z}.
\ee
In order to express the right-hand side in terms of $G_k$, we write it as
\be\notag
\left\langle
\sum_s 
p_s(Z_1,Z_2) 
e^{
- e^{-x}  F(Z_1,Z_2)
}
e^{
- \sum_i e^{-x_i} A_i Z_i
}
\right\rangle_{\mathbb{U},Z}
\Bigg|_{x_i\rightarrow x}.
\ee
We see that in the first two factors we can make the replacement
\be
Z_i \rightarrow - \f{1}{A_i} \f{\partial}{\partial e^{-x_i}} = \f{e^{x_i}}{A_i}\f{\partial}{\partial x_i}.
\ee
Then we can average over the $Z_i$, giving
\begin{align} \notag
G_{k+1}(x) \simeq
& \bigg\langle
\sum_s 
p_s\lf A_1^{-1}e^{x_1}\partial_1,A_2^{-1}e^{x_2}\partial_2\ri 
\\ \notag
& \times \exp \lf {-e^{-x} F(A_1^{-1}e^{x_1}\partial_1,A_2^{-1}e^{x_2}\partial_2) }  \ri
\\
& \times G_k(x_1-\ln A_1)
G_k(x_2-\ln A_2)
\bigg\rangle_{\mathbb{U}} \bigg|_{x_i\rightarrow x}. 
\label{eq:Geqnfull}
\end{align}
This equation is in principle exact.
For the case of forced measurements we would have a similar equation,  without the sum over $s$ or the probability factor $p_s$.

When $x$ is far to the right of the front (${x-\ln \ztyp\gg 1}$), the quantity ${H_k=1-G_k}$ is exponentially small in ${x-\ln \ztyp}$, and we can linearize the equation in $H$. 
Exploiting  the symmetry between $Z_1$ and $Z_2$, this linearization gives
\begin{align} \notag
\f{1}{2} H_{k+1}(x) \simeq
&  \bigg\langle
\sum_s 
p_s\lf A_1^{-1}e^{x_1}\partial_1, 0 \ri 
\\ \notag
& \times \exp \lf {-e^{-x} F(A_1^{-1}e^{x_1}\partial_1,0) }  \ri
\\
& \times H_k(x_1-\ln A_1)
\bigg\rangle_{\mathbb{U}} \bigg|_{x_1\rightarrow x}. 
\label{eq:Heqnfull}
\end{align}
When $\theta=\theta_c$, this equation has the exact solution  (for all $x$)
$H_{k+1}(x) = H_k(x) = \exp (-x)$.\footnote{One can check this by using  Eq.~\ref{eq:eqnformptthetac}, 
together with the fact that the expansion of $F$ in its argument starts at quadratic order, and the fact that $p_s(Z,Z')$ is normalised for all values of $Z,Z'$ (which means that ${\sum_s \partial_Z p_s(Z,0)|_{Z=0}=0}$).}

Let us consider the entangled phase, where ${G\equiv \lim_{k\rightarrow \infty} G_k}$ has a nontrivial limit, satisfying Eq.~\ref{eq:Geqnfull} with $G_k=G_{k+1}=G$.
If we are very close to the phase transition then $\ztyp$ is very small, and there is an ``intermediate regime'' of negative $x$  where
\be\label{eq:intermediatexregime}
1 \ll |x| \ll |\ln \ztyp|.
\ee
Then, in addition to linearizing the equation in $H$ (as above), we may also treat $e^x$ as a small quantity.
Neglecting all terms of order $e^x$ in 
(\ref{eq:Heqnfull}) gives\footnote{Note that ${-e^{-x} F(A_1^{-1}e^{x_1}\partial_1,A_2^{-1}e^{x_2}\partial_2)}$ is of order $e^x$, because the Taylor expansion of $F$ starts at quadratic order.}
\be
H(x) = 2 \left\langle \sum_s p_s(0,0) H(x-\ln A_1) \right\rangle.
\ee
This equation was essentially solved in Sec.~\ref{sec:real_linear}.
We use the ansatz $H(x)\propto \exp (- \lambda x)$. 
Because here we are considering a solution that is stationary (in $k$),
the wave parameter $\lambda$ must be chosen so that the velocity $v_\theta(\lambda)$ vanishes, as discussed in Sec.~\ref{sec:real_scaling}.
This condition gives the solution 
discussed in the main text, with
\be\label{eq:logslope}
1+ \partial_x \ln H  =  \f{c \sqrt{\theta_c-\theta}}{\tan (\phi + c
\sqrt{\theta_c-\theta} \, x) }.
\ee
As discussed in the main text, a heuristic matching argument at the left-hand-side of the range  of $x$ under discussion allows $\ln \ztyp$ be expressed in terms of $\phi/c\sqrt{\theta_c-\theta}$. The key point is that the second term in (\ref{eq:logslope}) should become non-negligible (and positive) 
 as $x$ approaches $\ln \ztyp$,
and this matching requires the argument of the tangent to vanish close to the front.

Next we consider the value of $\phi$.
In the case of \textit{forced} measurements the value of $\phi$ can be fixed straightforwardly (if non-rigorously) by matching on the other side of the intermediate-$x$ range. 
For forced measurements, a formula like Eq.~\ref{eq:logslope} holds, but with the first term being $-1/2$ instead of $-1$. 
Therefore the slope in the intermediate regime is close to $-1/2$. On the other hand, it is easy to show (for example from the definition of the generating function) that for $x\gg 1$ the slope must approach $-1$. Therefore, for forced measurements, we argue that $\phi=\pi$, in order that, as $x$ approaches 0 from the left, the slope should decrease (by an amount much larger than $\sqrt{\theta_c-\theta}$) to match onto the appropriate solution for $x\gtrsim 0$.
 
However for real measurements we have 
${\partial \ln H\simeq- 1}$ both in the intermediate regime and for $x \gg 1$. Therefore we cannot use the same argument to fix $\phi$.

Instead, it seems plausible that $\phi$ can be fixed to $\pi/2$ by more careful considerations along the following lines. We have (we consider the limit $k\rightarrow \infty$) 
\be
G(x) = \langle \exp(-e^{-x}Z) \rangle.
\ee
Differentiating to obtain an expression for the slope $\partial_x\ln H$ discussed above, and defining 
$\tilde Z = e^{-x}Z$, gives
\be
1+ \partial_x \ln H =  \f{\langle 1- e^{-\tilde Z} - \tilde Z e^{-\tilde Z} \rangle}{1- \langle e^{-\tilde Z}\rangle}.\label{eq:slopemoments}
\ee
Both the numerator and denominator may in principle be expanded in moments of $Z$ to all orders. The key point is  that while the expansion of the denominator starts with the first moment, $\langle Z\rangle$, the expansion of the numerator starts with the second moment.

Directly averaging the recursion relation for $Z_k$, in the stationary regime where the moments are independent of $k$, gives
\be
\langle Z \rangle = 2\langle A\rangle \langle Z \rangle
+ B,
\ee
where $B$ is a sum of terms involving higher moments (and higher powers of the first moment). 
Since ${2\langle A\rangle-1}$ is of order ${(\theta_c-\theta)}$, 
this equation
strongly suggests that, close to the critical point, 
the higher moments are smaller than the first moment by a factor of order ${(\theta_c-\theta)}$:
\be
\langle Z^n\rangle /\langle Z\rangle  = O(\theta_c-\theta) \qquad \text{for $n \geq 2$}.
\ee
We have confirmed this numerically for $n=2$.
Applied to Eq.~\ref{eq:slopemoments}, this  scaling implies 
that, in the limit where $\theta_c-\theta$ tends to  zero, but for any fixed $x$,
\be
1  + \partial_x \ln H \simeq (\theta_c-\theta) f(x),\label{eq:slopecorrectionfrommoments}
\ee
where $f(x)$ is an unknown function of $x$. (We have also checked this relation numerically, using the expression in Eq.~\ref{eq:slopemoments}.)

Next we would like to compare Eq.~\ref{eq:slopecorrectionfrommoments} with 
Eq.~\ref{eq:logslope}. 
If in Eq.~\ref{eq:logslope} we fix a value of $x$, and then take the limit of small ${\theta_c-\theta}$, we obtain 
${1+\partial_x \ln H \simeq c\sqrt{\theta_c-\theta}/\tan(\phi)}$.
At first glance, consistency with (\ref{eq:slopecorrectionfrommoments}) requires ${1/\tan(\phi)}$ to vanish, allowing us to fix ${\phi=\pi/2}$. However, to justify this conclusion we would need to  quantify the size of corrections to 
(\ref{eq:logslope}) arising from the terms (involving  $e^x$)  that we neglected in Eq.~\ref{eq:Heqnfull}, in order to check that the two approximations
(\ref{eq:logslope}) and 
 (\ref{eq:slopecorrectionfrommoments})
have a nonvanishing region of overlap at small ${\theta_c-\theta}$.

\bibliography{reference.bib}

\end{document}